\documentstyle[12pt,graphicx]{ioplppt}
\DeclareSymbolFont{AMSb}{U}{msb}{m}{n}     % iopfts.sty produced a lot of
\DeclareSymbolFontAlphabet{\mathbb}{AMSb}  % error messages

\begin{document}

\jl{6}
\title{Spin Foam Quantization and Anomalies}[Spin Foam Quantization and 
Anomalies]
\author{Martin Bojowald and Alejandro Perez}
\address{Center for Gravitational
    Physics and Geometry, The Pennsylvania
    State University, 104 Davey Lab, University Park, PA 16802, USA\\
E-mail: {\tt bojowald@gravity.phys.psu.edu} and {\tt perez@gravity.psu.edu}}

\newcommand{\va}{}
\newcommand{\phys}{\ \ =_{ \left.\right. _{\left.\right._{\!%
\!\!\!\!\!\!\!\!\!\!\!\!phys}}} \ \ }
\newcommand{\id}{{\rm 1\:\!\!\! I}}

\newcommand{\vt}{\vartheta}
\newcommand{\vp}{\varphi}

\newcommand*{\R}{{\mathbb R}}
\newcommand*{\C}{{\mathbb C}}

\newtheorem{rem}{Remark}
\newtheorem{ex}{Example}

\begin{abstract}
 The most common spin foam models of gravity are widely believed to be
 discrete path integral quantizations of the Plebanski
 action. However, their derivation in present formulations is
 incomplete and lower dimensional simplex amplitudes are left open to
 choice. Since their large-spin behavior determines
 the convergence properties of the state-sum, this gap has to be
 closed before any reliable conclusion about finiteness can be
 reached. It is shown that these amplitudes are directly related to
 the path integral measure and can in principle be derived from
 it, requiring detailed knowledge of the constraint algebra and
 gauge fixing.
% of its first class part which in the
% case of gravity generates space-time diffeomorphisms. 
% It has been
% suggested that the discretization of space-time in a spin foam model
% breaks the diffeomorphism gauge without introducing an explicit gauge
% fixing. Here we show that 
 In a related manner, minimal requirements of background
 independence
%--which are reminiscent of cylindrical consistency in
% loop quantum gravity--
 provide non trivial restrictions on the form of
 an anomaly free measure. Many models in the literature do not satisfy
 these requirements.
% Moreover, we show that an anomaly free model will
% necessarily contain divergent amplitudes that could be interpreted as
% due to infinite contributions of gauge equivalent
% configurations. Exploring these issues we come across a 
 A simple model satisfying the above consistency requirements is
 presented which can be thought of as a spin foam quantization of the
 Husain--Kucha\v{r} model.
\end{abstract}

\pacs{0460P}

\section{Introduction}

In recent years, spin foam models have been established as possible
candidates for a quantum theory of gravity (for recent reviews see
\cite{review,ori2}).  They are commonly viewed as covariant (path
integral) versions of a canonical quantization and in fact share some
features of quantum geometry (in the case of 2+1 a precise relation
can be established \cite{Noui:2004iy}). As a discretized path integral
they can be derived from Plebanski's action \cite{a16} which is a
formulation of general relativity as a constrained $BF$-theory
\cite{pleb}. Being path integrals of a gauge theory, they have to deal
with the anomaly issue: the path integral measure has to be invariant
under transformations generated by the constraints. Sometimes it is
claimed that a covariant quantization avoids the issue of anomalies
which plagued canonical approaches for a long time (but see
\cite{th2}).  However, it is well known that there is also an anomaly
problem in path integral quantizations which in spin foam
quantizations has just been ignored in most of the existing literature
(see, however, \cite{fre7} for a recent paper which discusses this
issue independently in the example of 3-dimensional $BF$-theory). A
complete analysis would require an understanding of the continuum
limit which has not yet been developed sufficiently.  Still, we will
see that it is possible to shed light on the problem and to derive
conditions for the amplitudes involved in the definition of a spin
foam model. To that end we look at the problem from two perspectives.
First, we view the discretized version, which is obtained by fixing a
space-time triangulation, as a regularization of the path integral, in
particular its measure, which would result in the continuum
limit. Second, we analyze the restrictions on the spin foam measure
imposed by background independence directly at the discrete level.

To explain the part of the anomaly problem studied here we first
recall the situation of standard path integrals: There is a
prescription which results in a unique measure (up to a constant
factor) which is at least formally invariant. This is usually the
obvious measure which integrates over all canonical coordinates
with constant weight function. After removing the regulator the
measure might not be invariant giving rise to quantum anomalies.
In any case, it is necessary to use the invariant measure for the
regularized version; otherwise the gauge symmetries are broken
explicitly and the results are unphysical. Since the formally
invariant measure is obvious in most cases, the standard term
`anomaly' only refers to the second issue, namely whether or not
the measure will remain invariant after removing the regulator.
\footnote{We warn the reader that
  anomalies in quantum field theories can even refer to quantities
  having observable consequences, which has no relation to the
  discussion here. We are only concerned with anomalies of gauge
  theories which must be absent for physical viability.}

In the case of spin foam models, the situation is more involved.
First, the constraint algebra is mixed and not closed which will be
seen to lead to an additional function in the measure which has not
been taken into account previously. Secondly, the space-time
discretization obscures the role of the measure and the meaning of
invariance in this context. Therefore, even analyzing the formal
invariance of a measure requires new techniques which will be provided
in the present paper. The usual anomaly problem, which analyzes the
invariance after the regulator is removed, will not be touched here
since the continuum limit is not understood. We will, however, see
that already a formally invariant measure, which is a necessary
prerequisite for an anomaly-free continuum measure, puts strong
restrictions on amplitudes in the spin foam model. Our definition of a
formally invariant {\em discrete\/} measure, i.e., amplitudes
associated with lower dimensional simplices, is that it must descend
from the formally invariant continuum measure along the lines of the
spin foam discretization. This does not imply that the continuum
formulation is seen as fundamental such that the discretization is
introduced only for, e.g., computational purposes. Rather, we view the
discrete theory as basic and use the continuum formulation as a tool
to derive conditions since it must at least be approximated in
certain, semiclassical regimes. This viewpoint concerning the lower
dimensional amplitudes and the continuum measure is the same one as
usually assumed for the relation between the vertex amplitude and the
continuum action.  Provided the required calculations are feasible,
this will also fix the formally invariant measure uniquely up to a
constant factor, which translates in conditions for the spin foam
amplitudes.

An immediate question in the context of anomalies is whether or
not a spin foam state sum can be finite.\footnote{A partition
function does not need to be finite since it usually drops out of
observable expressions due to normalization. Nevertheless, in the
context of gravity finite state sum models have been discussed,
for which both the partition function and transition amplitudes
are finite. In gravity, the viewpoint is different from common
field theories since state sums are used directly to compute
projection operators to the physical Hilbert space, and not only
to compute observables.} The aim of the present paper is to
devise methods for checking a spin foam quantization for formal
anomalies, thus obtaining conditions for the lower dimensional
amplitudes.  While there is general agreement on the vertex
($4$-simplex) amplitude, which can be viewed as representing the
exponentiated action, there are no clear-cut arguments as to which
lower dimensional simplex amplitudes should be used\footnote{In
the case of the Barrett--Crane model the normalization that yields
finite amplitudes is naturally selected in the context of the
group field theory (GFT) formulation \cite{reis1,a10}; however, no
clear connection with the formal path integral has not been found
yet.}; in the literature, it is largely regarded as being open to
choice, maybe constrained by semi-classical issues. This problem
is particularly pressing because the question of whether or not a
model is finite hinges on the asymptotic behavior of these
amplitudes. 
%In fact, we will see that these lower dimensional
%simplex amplitudes represent a discretized version of some part of
%the path integral measure and can be {\em derived\/} from it.
%Choosing different amplitudes is equivalent to inserting an
%arbitrary function into the path integral; then it is very easy to
%get a finite model by introducing a suppression of the measure
%along the orbits. However, such an anomalous model has to be
%dismissed as unphysical.

Note that our criterion is formulated from the perspective that the
fundamental theory is intrinsically discrete.
No matter how the approach to a continuum description is
performed --via a limit or as a coarse-grained approximation-- gauge
degrees of freedom have to be removed which is only possible with an
invariant measure. It is sometimes argued that
finiteness arises because ultraviolet or infrared divergences are
regularized by quantum gravitational effects like a minimal length
scale (this does in fact occur in canonical quantizations
\cite{QSDV}). From our point of view, however, this is not tenable
since the anomaly issue is completely unrelated to ultraviolet or
infrared divergences.

As a consequence of an invariant measure, a standard path integral
quantization cannot lead to finite results without gauge fixing if
gauge orbits do not have finite volume (which is to be expected for
gravity; to avoid confusion we emphasize here that we are mainly
concerned with the diffeomorphism constraints, not with an $SO(4)$ or
$SL(2,\C)$ Gauss constraint).\footnote{Sometimes it is argued that the
breaking of active space-time diffeomorphism invariance by a
triangulation allows finite results. However, in this argument one has
to invoke the usual equivalence of active and passive diffeomorphisms,
but the very triangulation which breaks active diffeomorphisms also
breaks the correspondence between active and passive transformations.}
A simple model where this can be illustrated is $2+1$ dimensional
$BF$-theory. This theory is equivalent to $2+1$ dimensional
gravity for non-degenerate triads and thus has the same gauge
orbit structure at least on the constraint surface. It turns out,
and is commonly accepted, that the spin foam amplitude for $2+1$
dimensional $BF$-theory is infinite in accordance with the
expectation from path integrals. The discrete symmetries of the
simplicial action can be explicitly analyzed \cite{fre7} and
directly linked with the triangulation independence of the spin
foam model.  An interesting case is $2+1$ gravity with
cosmological constant $\Lambda$.  In this case the action can be
written as that of a $Spin(4)$ Chern-Simons theory whose level $k$
is given by $k=4\pi/\sqrt{\Lambda}$ (see \cite{baez5} and
references therein). A path integral quantization of this theory
leads to the Turaev-Viro model defined in terms of a quantum group
$SU_q(2)$ for $q$ a root of unity related to the cosmological
constant by $q={\rm exp}[2\pi \i/(k+2)]$. Transition amplitudes
turn out to be finite. Although this is often interpreted as a
consequence of the infrared cut-off introduced by the quantum
deformation, from our viewpoint this is a consequence of the
compactness of the gauge group $Spin(4)$.

It is not clear however how to generalize this intuition to four
dimensions. The gauge properties of $BF$-theory in four dimensions
are very similar to its $3$-dimensional relative. In particular,
divergences in the path integral \cite{crane0} can also be traced
back to infinite volume factors coming from the topological gauge
symmetry. If we concentrate on the spin foam models for four
dimensional gravity that are obtained from an implementation of
constraints on the $BF$ amplitudes (such as
Reisenberger\cite{reis4} or the Barrett--Crane models
\cite{BC2,BC1}) the topological gauge symmetry is manifestly
broken by the implementation of the constraints. As a result, it
was debated whether the remnant gauge symmetries would produce
diverging spin foam amplitudes or rather contain ``finite volume''
gauge orbits. Here we show that minimal requirements of background
independence imply the existence of divergences and rule out the
finite normalizations proposed in the literature
\cite{a7,a2,a22,baez1}.

%On general grounds, one would expect a discretization of gravity to
%break the diffeomorphism invariance, but this by itself does not
%guarantee that the resulting model will be finite. In fact, if the
%discretization becomes finer and finer, thus approaching the continuum
%limit, the discretization should become more and more invariant. This
%justifies our condition for a formally anomaly-free discrete measure
%as being derived from the invariant continuum measure.

In the following section, we will introduce a finite dimensional
toy model which illustrates the steps of a spin foam quantization
mimicking $BF$-theory with additional constraints. In
Section~\ref{measure} we discuss the definition of the (formal)
path integral for constrained systems. In Section~\ref{BF} we
revisit the spin foam quantization of $BF$-theory in three
dimensions to introduce notation and review the gauge analysis of
the discrete theory performed in \cite{fre7}.  In Section~\ref{GR}
we discuss the definition of the correct path integral measure for
4-dimensional spin foam models defined as constrained BF state
sums.  We look at the problem from the passive and active
diffeomorphism perspectives.  In the first case, we reformulate
$BF$-theory in a way which makes the relation between the path
integral measure and the face amplitude obvious. This provides us
with a recipe for computing the large spin behavior of amplitudes
in spin foams for gravity discussed in Subsection~\ref{GRe}. In
Subsection~\ref{DC} we discuss the restrictions imposed on the
form of the measure by background independence in the active
picture. In the case of the Barrett-Crane model, we show that
various normalizations proposed in the literature do not satisfy
these requirements and should be regarded as anomalous. This
includes the finite normalization introduced in \cite{a10}. In
Subsection~\ref{te} we define a simple model satisfying those {
background independence} requirements.  The latter can be thought
of as the spin foam quantization of the Husain--Kucha\v{r} model.

\section{A toy model}\label{TM}

To illustrate the importance of choosing the correct measure in spin
foam models we first discuss a simple toy model with a finite number
of degrees of freedom. It incorporates the essential steps of a spin
foam quantization of Plebanski's action for gravity, which are a field
discretization and the solution of a constraint for Lagrange
multipliers. Being a system with a finite number of degrees of freedom,
the continuum limit cannot be modeled. However, as discussed before,
the anomaly issue already requires the correct treatment of the
regularization before the continuum limit is taken, which will be
illustrated here.

\subsection{Definition and evaluation}

The action of the model is given by
\begin{equation}
 S=\int\d t(\dot{q}_1p_1+\dot{q}_2p_2+\lambda_1q_1+\lambda_2
 q_2+\xi(\lambda_1-\lambda_2))
\end{equation}
which has two constrained degrees of freedom\footnote{The system
    is not constrained completely since after solving the constraint
    obtained after varying $\xi$ there is only one constraint for two
    degrees of freedom. The remaining constraint requires $q_1+q_2$ to
    be zero, but otherwise $q_1=-q_2$ is free. Since the usual kinetic
    term is missing, the dynamics then requires $q_1$ to be constant,
    which we will also see later in the path integral quantization.}
    $(q_1,q_2)$, which we assume to live on a circle, with conjugate
    momenta $(p_1,p_2)$ and three Lagrange multipliers $\lambda_1$,
    $\lambda_2$ and $\xi$. Compared with $BF$-theory,
    $(p_1,p_2,\lambda_1,\lambda_2)$ represents the components of the
    field $B$ which contains both physical degrees of freedom and
    Lagrange multipliers. If we set $\xi=0$ resulting in the action
\[
 S|_{\xi=0}=\int\d t(\dot{q}_1p_1+\dot{q}_2p_2+\lambda_1q_1+\lambda_2
 q_2)\,,
\]
the theory is constrained completely, i.e., both $q_1$ and $q_2$ must
be zero. There are no degrees of freedom in this case.
With unrestricted $\xi$, however, the two original Lagrange
multipliers are constrained which restores one degree of freedom:
$\lambda_1$ has to equal $\lambda_2$ and thus only $q_1+q_2$ has to be
zero whereas the difference is free, which can easily be seen by
solving the $\xi$-constraint explicitly:
\[
 S=\int\d t(\dot{q}_1p_1+\dot{q}_2p_2+\lambda_1(q_1+q_2))\,.
\]
This feature mimics the
transition from $BF$-theory to gravity where also additional
constraints (the simplicity constraints) reduce the freedom of
original Lagrange multipliers of $BF$-theory and thereby introduce
local degrees of freedom.

A spin foam quantization proceeds by quantizing the simple theory
whose discretized state sum can be computed explicitly and
incorporating the additional constraints at the state sum level. The
simple theory (the analog of $BF$-theory) here is $S|_{\xi=0}$ with
path integral
\begin{eqnarray}\label{Z0}
\fl Z_0&=\int{\cal D}^2q{\cal D}^2p{\cal D}^2\lambda\e^{\i S|_{\xi=0}}
% \nonumber\\ &
= \int{\cal D}^2q{\cal D}^2p\exp\left(\i\smallint\d
 t(\dot{q}_1p_1+\dot{q}_2p_2)\right) \delta(q_1)\delta(q_2)= \int{\cal
 D}^2p
\end{eqnarray}
with ${\cal D}^2q:={\cal D}q_1{\cal D}q_2$. The result is certainly
infinite since we are dealing with an unfixed gauge theory. In this
case a gauge fixing is simple, but we do not do this because we want
to understand the role of a field discretization and the multiplier
constraints in this respect.

The path integral for $S$, the analog of gravity, in this case can
also be obtained explicitly:
\begin{eqnarray}
\fl Z = \int{\cal D}^2q{\cal D}^2p{\cal D}^2\lambda{\cal D}\xi\e^{\i
 S}
%\\\nonumber \lo
= \int{\cal D}^2q{\cal D}^2p{\cal D}\lambda_1\exp\left(\i\smallint\d
 t(\dot{q}_1p_1+\dot{q}_2p_2+\lambda_1(q_1+q_2))\right) \label{Z}\\
 \fl = \int{\cal
 D}^2q{\cal D}^2p\exp\left(\i\smallint\d t(\dot{q}_1p_1+\dot{q}_2p_2)\right)
 \delta(q_1+q_2)
%\nonumber\\ \lo
= \int{\cal D}q_1{\cal D}\Delta p\exp\left(\i\smallint\d
 t\dot{q}_1\Delta p\right)\int{\cal D}p' \nonumber
\end{eqnarray}
with $\Delta p:=p_1-p_2$ and $p'=p_1+p_2$. Computing the remaining
integrations we obtain
\begin{equation}
 Z=\int{\cal D}q_1\delta(\dot{q}_1)\int{\cal D}p'=
 \delta(q_1^{(0)}-q_1^{(1)})\int{\cal D}p'\,.
\end{equation}
Here, $q_1^{(0)}$ and $q_1^{(1)}$ represent the initial and the final
value of $q_1$ which are constrained to be equal but free
otherwise. Due to the fact that we have only one remaining gauge
symmetry after incorporating the $\xi$-constraint, we only have one
infinite integral left rather than two in (\ref{Z0}).

In finite spin foam models one solves the multiplier constraint
$\lambda_1-\lambda_2=0$ at the discretized level and, in some cases,
obtains a finite result even {\em without\/} fixing the remaining
gauge freedom \cite{a1}. A spin foam quantization, however, also
involves a discretization of space-time which, as already mentioned,
is not realized in this finite dimensional model. Still, it is worth
checking what effects a field discretization itself can have; effects
of the space-time discretization will be discussed later. Therefore,
we now discretize $\lambda_1$ and $\lambda_2$ which are analogous to
$B$-field components (we could also discretize the remaining
components $p_1$ and $p_2$, without changing our results). In analogy
to a spin foam quantization we do this by writing the integral
representation of the delta function
$\delta(q_1)=(2\pi)^{-1}\int\d\lambda_1\e^{\i\lambda_1q_1}$ as a sum
$\delta(q_1)=(2\pi)^{-1}\sum_{n_1}\e^{\i n_1q_1}$.  The path integral
(\ref{Z0}) (absorbing constant factors in the measure) then becomes
\begin{equation}\label{Z0disc}
 Z_0=\int{\cal D}^2q{\cal D}^2p\sum_{\{n_1\},\{n_2\}}\exp\left(\i\smallint\d
 t(\dot{q}_1p_1+\dot{q}_2p_2+n_1q_1+n_2 q_2)\right)
\end{equation}
where the summation index $\{n\}$ indicates that $n$ is not a single
number but a function of time, and we are summing over the values at
fixed times individually (i.e., this is a discrete analog of the path
integral).

At the discrete level the $\xi$-constraint implies $n_1=n_2$ and we
must only sum over those pairs of integers fulfilling this condition
in order to obtain a quantization for $S$ (analogously to summing only
over simple representations in a spin foam quantization):
\begin{equation} \label{Zdisc}
 Z=\int{\cal D}^2q{\cal D}^2p\sum_{\{n_1\}}\exp\left(\i\smallint\d
 t(\dot{q}_1p_1+\dot{q}_2p_2+n_1(q_1+q_2))\right)\,.
\end{equation}
This is the analog of (\ref{Z}) and its value is, of course, the
same as in the calculation with continuous $\lambda_1$. In
particular, it is infinite. As anticipated, the field
discretization and the spin foam like quantization could not take
care of the gauge orbit divergence. The divergent integral of
(\ref{Z}) has just been
  replaced by a divergent sum in (\ref{Zdisc}).

\subsection{Modifying the measure}\label{TMT}

In spin foam quantizations the issue of convergence hinges on the
choice of lower dimensional simplex amplitudes, which can be
considered as functions of some components of the discretized
$B$-field.  In our model, however, we do not have any free function
available since the path integral result is unique.
% As we will discuss
%later, the lower dimensional simplex amplitudes of spin foams also are
%fixed uniquely (up to different discretization choices), but have not
%been determined yet. 
To include such a function we write our result in the spin foam form
(there is still a $p$-integral because we chose not to discretize $p$)
$Z=\int{\cal D}^2p\sum_{\{n_1\}}V(p_1,p_2,n_1)$ where
\[
 V(p_1,p_2,n_1):=\int{\cal D}^2q\exp\left(\i\smallint\d
 t(\dot{q}_1p_1+\dot{q}_2p_2+n_1(q_1+q_2))\right)
\]
is the vertex amplitude (analogous to the integration over connections
of the discretized $\e^{\i S}$). A model of lower dimensional amplitudes
can now be included by simply inserting a new function $A(n_1)$ into
$Z$ (more generally, $A$ could also depend on $p$):
\[
 Z=\int{\cal D}^2p\sum_{\{n_1\}}A(n_1)V(p_1,p_2,n_1)\,.
\]
Our derivation shows that the face amplitude $A(n_1)$ is fixed and
constant, but let us see what a different function would imply. For
illustrative purposes, we choose $A(n_1)=(2n_1+1)^{-2}$ which is
finite for all integer $n_1$. Now it is easy to see that
\begin{eqnarray*}
\fl Z' = \int{\cal D}^2p\sum_{\{n_1\}}(2n_1+1)^{-2}V(p_1,p_2,n_1)\\
 \lo= \int{\cal
 D}^2q {\cal D}^2p\sum_{\{n_1\}}(2n_1+1)^{-2}\exp\left(\i\smallint\d
 t(\dot{q}_1p_1+\dot{q}_2p_2+n_1(q_1+q_2))\right)\\
 \lo= \int{\cal
 D}^2q {\cal D}^2p \exp\left(\i\smallint\d
 t(\dot{q}_1p_1+\dot{q}_2p_2+V_{{\rm eff}}(q_1,q_2))\right)
\end{eqnarray*}
{\em is\/} finite, where we have the effective potential
\begin{equation}
 V_{{\rm eff}}(q_1,q_2)=\log(q_1+q_2-\pi)-\case{1}{2}(q_1+q_2)\,.
\end{equation}
(We used the Fourier series $\sum_k(2k+1)^{-2}\e^{\i k\phi}=
-\frac{\pi}{4}(\phi-\pi)\e^{-\i\phi/2}$ for $0\leq x<2\pi$ and extended
with $2\pi$-periodicity.) In fact, this is an ordinary path integral
for a system of two degrees of freedom in an effective potential
$V_{{\rm eff}}$ {\em without\/} constraints. We now have to decide if
this finite result makes sense and can tell us anything about the
original system. The answer is clearly {\em negative}: The role of the
effective potential is completely unclear, and it has nothing to do
with the original system. Originally, $q_1$ and $-q_2$ have to equal
each other but are free otherwise, whereas in the modified system they
are independent but subject to motion in a potential. Furthermore, the
kind of modification, e.g.\ the form of the potential, depends on the
face amplitude which has no distinguished form other than $A(n_1)=1$
which follows from the invariant measure. In conclusion, a finite path
integral for an unfixed system with constraints cannot be trusted. (It
cannot even be regarded as an approximation since the measure is not
just a smeared version of a $\delta$-function with support on the
constraint surface. The effective potential is singular on a
submanifold of the configuration space, but this does not happen at
the constraint surface $q_1+q_2=0$, but at $q_1+q_2=\pi$.) In fact,
introducing a non-constant amplitude is nothing but introducing
an arbitrary function $A(\lambda_1)$ into the path integral which {\em
  breaks the invariance of the measure}. (Note that $\lambda_1$ serves
as a Lagrange multiplier and thus its conjugate momentum
$p_{\lambda_1}$ is implicitly constrained to be zero. The gauge
freedom generated by this constraint is broken by introducing an
arbitrary function of $\lambda_1$ into the measure. Consequently, the
multiplier $\lambda_1$ is no longer completely free which also affects
the remaining gauge freedom.) This explains why we get a finite result
with independent $q_1$, $q_2$; and it also demonstrates that here a
finite model is {\em anomalous}. As discussed in the Introduction, the
space-time discretization, which is not modeled here, presents a
possible rescue for finite spin foam models. To check this, we need
more general methods which will be introduced in what follows.

\section{General discussion}
\label{measure}

Since our model incorporated some of the essential steps of a spin
foam quantization of gravity, it suggests that the same conclusions
regarding the choice of amplitudes hold true in this more complicated
case. In this section we discuss the continuous path integral and the
correct measure in the presence of second class constraints, which
will be necessary to derive the anomaly-free amplitudes.

The characteristic feature of the gravitational action which is
commonly used for a spin foam quantization is the presence of a
constraint which restricts the allowed values of Lagrange multipliers
appearing in a simpler action. 
%We illustrated this property in the previous
%toy model where the importance of an invariant path integral measure has
%been seen explicitly. 
To find the correct measure it is not sufficient
to work solely in a Lagrangian formulation; in particular it is
essential to understand the structure of the constraint algebra which
can only be achieved in a Hamiltonian analysis. The constraint algebra
in this context is always mixed (i.e.\ neither purely first class nor
purely second class) and rather complicated. There are always the
usual diffeomorphism constraints of gravity which must form a suitable
first class sub-algebra, but in this particular formulation there is
also a second class contribution: a constraint which restricts the
multipliers of other constraints must be second class. Despite first
appearance, even in the toy model the additional constraint is second
class. Although the constraints $C_1=q_1$, $C_2=q_2$ and
$C_3=\lambda_1-\lambda_2$ Poisson commute, one has to take into
account that in this form they are constraints on a {\em
non}-symplectic Poisson manifold with coordinates
$(q_1,p_1;q_2,p_2;\lambda_1,\lambda_2)$ where the standard definitions
of Dirac's classification do not apply (see \cite{brackets} for a
discussion and generalized definitions). One can easily introduce an
equivalent constrained system which has constraints on a symplectic
manifold by adding the momenta $\pi_1$ and $\pi_2$ conjugate to the
restricted multipliers $\lambda_1$ and $\lambda_2$, together with the
constraints $C_4=\pi_1$, $C_5=\pi_2$. The constraints $C_I$,
$I=1,\ldots,5$ are then defined on a symplectic manifold and now it is
obvious that $C_3$ does not commute with all constraints. In fact
$C_3$ and $C_4-C_5$ form a second class sub-algebra,
$\{C_3,C_4-C_5\}=2$, whereas $C_1$, $C_2$ and $C_4+C_5$ are first
class.

%The presence of second class constraints requires a special treatment
%when deriving the correct measure. It is not sufficient simply to
%include an integration over the multipliers since this leaves open an
%arbitrary function. 
In the absence of constraints the invariant path
integral measure is given by the determinant of the symplectic form
which leads to ${\cal D}q{\cal D}p$ for canonical coordinates
$(q,p)$. A similar treatment is not possible for the multiplier
integration since multipliers form a Lagrangian sub-manifold of the
extended phase space such that the determinant of their symplectic
structure would be zero. The measure is well-defined after solving the
second class constraints (and turning the first class constraints into
second class ones by fixing the gauge), which leads to the symplectic
structure following from the Dirac bracket. For completeness, we will
next show how to derive the correct treatment of the multiplier
measure by requiring that after solving the constraints in the
integral we obtain the determinant of the Dirac symplectic structure
\cite{Fad}.

We start with a system with $2n$ coordinates $x_i$, $i=1,\ldots,2n$ on
a symplectic phase space $(M,\omega)$ and $m$ second class constraints
$C_I$, $I=1,\ldots,m$. (There might be additional, first class
constraints which are not relevant for this section. They can either
be gauge fixed and included in the constraints $C_I$ or be left for
later treatment, e.g.\ factoring out the volume of their gauge
orbits. The second possibility is particularly interesting here since
a gauge fixing is sometimes claimed to be unnecessary for spin foam
models of gravity.)  The path integral (where the constraint part has
been split off the action $S=S_0+\int\sum_I\lambda^IC_I$) then is
\[
 Z=\int{\cal D}^{2n}x\,\sqrt{\det\omega}\,{\cal D}^m\lambda
 \,\mu(x) \exp\left(\i\smallint\sum_I\lambda^IC_I\right) \exp(\i S_0)
\]
with a function $\mu(x)$ for the multiplier measure which will be
determined shortly. This function must not depend on the $\lambda^I$
because otherwise the multiplier integration would not yield
$\delta$-functions of the constraints. With a $\lambda$-independent
$\mu$ we can perform the $\lambda$ integrations explicitly and obtain
\[
 Z=\int{\cal D}^{2n}x\,\sqrt{\det\omega}\,\mu(x)\prod_I
 \delta(C_I) \exp(\i S_0)\,.
\]

To proceed further we transform from the coordinates $x_i$,
$i=1,\ldots,2n$ to coordinates $(y_{\alpha};C_I)$ with
$\alpha=1,\ldots, 2n-m$, $I=1,\ldots,m$ (assuming that the constraints
are regular and irreducible such that they can be used as local
coordinates on $M$). To find the Jacobian of this transformation we
use the fact that locally the symplectic manifold $(M,\omega)$ can be
represented as $(M,\omega)\cong(R,\omega_D) \times_R (P,\Pi_P^{-1})$
using the following notation. The symplectic manifold $(R,\omega_D)$
is the constraint surface $R\subset M$ defined by $C_I=0$,
$I=1,\ldots,m$, endowed with the Dirac symplectic structure
$\omega_D$. The manifold $P$ is given by the image of a neighborhood
of a point in $R$ under the functions $C_I\colon M\to\R$ (i.e.\ $P$ is
a neighborhood of $0$ in $\R^m$; for our purposes it is sufficient to
know $P$ only locally) and coordinatized by $(C_I)$,
$I=1,\ldots,m$. If the constraint algebra is closed,\footnote{i.e.,
the right hand side consists of functions of the constraints and thus
is constant on the constraint surface} $P$ can be defined globally and
is a Poisson manifold with Poisson tensor $\Pi_P$ defined by
$\Pi_P(\d C_I,\d C_J):=\{C_I,C_J\}$ where the bracket on the right
hand side is computed using the symplectic structure $\omega$ on $M$
\cite{ASS}. For second class constraints the inverse of $\Pi_P$ exists
and $(P,\Pi_P^{-1})$ is a symplectic manifold. If the constraint
algebra is not closed, the Poisson tensor $\Pi_P$ depends not only on
the coordinates $C_I$ of $P$, but also on the coordinates of $R$ such
that $(P,\Pi_P)$ as a symplectic manifold depends on the point in $R$
chosen for its definition. The right component of the product
decomposition of $M$ then depends on a point in the left component,
which is indicated by the subscript $R$ of the symbol $\times$. That
the decomposition is valid locally can be shown using the methods of
\cite{brackets} where it has been proven for a closed algebra.

Here we use this local decomposition to factor the original symplectic
structure $\omega=\omega_D\otimes\Pi_P^{-1}$ which allows us to
perform the coordinate transformation in the last integral,
\begin{eqnarray*}
 \fl Z = \int{\cal D}^{2n-m}y{\cal D}^mC
 \sqrt{\det\omega_D/\det(\{C_I,C_J\})}\,
 \mu(y,C)\prod_I\delta(C_I) \exp(\i S_0)\\
 \lo= \int{\cal D}^{2n-m}y \sqrt{\det\omega_D}
 \,\mu(y,0)/\sqrt{\det(\{C_I,C_J\})} \exp(\i S_0)\,.
\end{eqnarray*}
Since $y_{\alpha}$ are coordinates on the constraint surface, their
path integral measure must be given by the Dirac symplectic
structure. If there is any other non-constant function besides the
exponential of the action, the measure will not be invariant and the
path integral will be anomalous. Therefore, the function $\mu(x)$ has
to be $\sqrt{\det(\{C_I,C_J\})}$ which fixes the free function in the
original integral (strictly speaking, $\mu$ is only fixed on the
constraint surface). In the presence of second class constraints,
therefore, the path integral to start with is
\begin{equation}\label{Zsec}
\fl Z=\int{\cal D}^{2n}x{\cal D}^m\lambda \sqrt{\det\omega}
  \sqrt{\det(\{C_I,C_J\})}
  \exp\left(\i\smallint\sum_I\lambda^IC_I\right) \exp(\i S_0)
\end{equation}
which requires a detailed knowledge of the constraint algebra. Note
that the determinant of the constraint brackets also appears in this
form when first class constraints $D_{\alpha}$ are gauge fixed a la
Faddeev--Popov, where the other half of the second class constraints
$C_I$ are gauge fixing conditions $f_{\beta}$:\footnote{In a covariant
formulation one usually chooses a gauge fixing functional depending
on all components of the fields, which would require the use of the
extended phase space in a canonical picture. For pure first class
constraints it is expected that a covariant gauge fixing would be
better suited to the spin foam approach; an example can be found in
\cite{fre7} and in Section \ref{BF}.} In this case the
measure contains a function $\det(\partial
f_{\alpha}/\partial\delta_{\beta}|_{\delta=0})$ where $f_{\alpha}$ is
a gauge fixing condition for the first class constraint $D_{\alpha}$
and $\delta_{\alpha}$ its gauge parameter. Thus, $\partial
f_{\alpha}/\partial\delta_{\beta}|_{\delta=0}=
\{C_{m/2+\alpha},C_{\beta}\}|_{C=0}$ (we assume that the second class
constraints are arranged in such a way that the first $m/2$ are the
original first class constraints, $C_I=D_I$ for $1\leq I\leq m/2$, and
the rest are the gauge fixing conditions $C_I=f_I$ for $m/2<I\leq m$)
and
\begin{equation}\label{Zsec1}
 \fl\det(\partial f_I/\partial\delta_J|_{\delta=0})=
 \det\{C_I,C_J\}|_{C=0;1\leq I\leq m/2;m/2<I\leq
 m}=\sqrt{\det\{C_I,C_J\}}|_{C=0}\,.
\end{equation}

In existing spin foam quantizations the correct factor for the
multiplier integration has not been taken care of; instead any
multiplier has been associated simply with a measure ${\cal D}\lambda$
without justification. 
%While this simplifies the analysis and avoids a
%discussion of the constraint algebra, in general it introduces
%anomalies and is not permissible; it also needs to be included if some
%second class subalgebra is treated before solving first class
%constraints (without discussing gauge fixing). 
Note that the
additional factor can be ignored when it is constant on the constraint
surface, which is always the case for a closed constraint algebra. In
particular, our toy model has a closed algebra and so our treatment
was correct even though we ignored the additional factor in the
measure. For more complicated systems including gravity in the
Plebanski formulation, however, this is not expected to be the case.

%In this section we have recalled the correct choice of measure in
%the continuum path integral in the presence of second class constraints.
%In the remainder we will see how it can be built into the discrete spin
%foam version and how it affects the amplitudes.

\section{$BF$-theory}\label{BF}

$BF$-theory is important in what follows because it appears as an
intermediate step in the definition of the gravitational spin foam
model. In this section we use it also to illustrate the possible role
played by the space-time discretization in the context of anomalies.

The action of $BF$-theory is given by
\begin{equation}\label{bfaction} S(B,A)=\int
\limits_{\cal M} \Tr(B\wedge F(A)),
\end{equation}
where the field $A$ corresponds to a connection on a principal bundle
with compact structure group $G$ (which will later be taken to be
$SU(2)$, which gives 3-dimensional Riemannian gravity) and the field
$B$ is a Lie algebra valued $1$-form. The local symmetries of the
action correspond to the internal gauge transformations
\begin{equation}\label{gauge1}
\delta B = \left[B,\omega \right], \ \ \ \ \ \ \ \ \ \delta A =
d_{A} \omega,
\end{equation}
for $\omega$ a Lie algebra valued scalar field where $d_{\va A}$
denotes the covariant exterior derivative, and `triad translations'
\begin{equation}\label{gauge2}
\delta B = d_{\va A} \eta, \ \ \ \ \ \ \ \ \ \delta A = 0,
\end{equation}
where $\eta$ is a Lie algebra valued function. The first invariance is
manifest from the form of the action, while the second is a
consequence of the Bianchi identity, $d_{\va A}F=0$. If one writes the
theory in the Hamiltonian formulation, one observes that the previous
symmetries are gauge symmetries in the Dirac sense, i.e., they are
generated by the Poisson bracket with the corresponding first class
constraints; there are no second class constraints.

Moreover the number of constraints equals the number of
configuration variables of the phase space, which implies the
theory can only have global degrees of freedom. This can be
checked directly by writing down the equations of motion
%\begin{equation}\label{soluto}
%F(A)=\d_{\va A}A=0, \ \ \ \ \ \ \ \ \ \d_{\va A}B=0.
%\end{equation}
$F(A)=\d_{\va A}A=0$, $\d_{\va A}B=0$.
The first equation is solved by flat connections which are locally
gauge. The solutions of the second equation are also locally
gauge, once the flatness condition ($F(A)=0$) holds, as any closed
form is locally exact.\footnote{\label{remark}
As is well known, one can easily check that the infinitesimal
diffeomorphism gauge action $\delta B={\cal L}_v B$, and $\delta
A={\cal L}_v A$, where ${\cal L}_v$ is the Lie derivative in the $v$
direction, is a combination of (\ref{gauge1}) and (\ref{gauge2}) for
$\omega=v^aA_a$ and $\eta_b=v^a B_{ab}$, respectively, acting on the
space of solutions.}
%, i.e. when (\ref{soluto}) holds.}

\subsection{Derivation of the spin foam model}\label{qbf}

To fix our notation, we will now discuss the spin foam quantization of
three-dimensional gravity, where ${\cal M}$ is a three-dimensional
manifold and $G=SU(2)$, and later mention necessary changes for four
dimensions.  The quantization of $BF$-theory is done by replacing the
manifold ${\cal M}$ with an arbitrary cellular decomposition
$\Delta$. We also need the notion of the associated dual 2-complex of
$\Delta$ denoted by ${\cal J}_{\Delta}$. The dual 2-complex ${\cal
J}_{\va \Delta}$ is a combinatorial object defined by a set of
vertices $v\in {\cal J}_{\va \Delta}$ (dual to 3-cells in $\Delta$)
edges $e\in {\cal J}_{\va \Delta}$ (dual to 2-cells in $\Delta$) and
faces $f\in {\cal J}_{\va \Delta}$ (dual to $1$-cells in $\Delta$).

The fields $B$ and $A$ have support on these discrete structures by
representing the $su(2)$-valued $1$-form $B$ as an assignment of a $B
\in {su(2)}$ to each $1$-cell in $\Delta$ the connection $A$ as an
assignment of a group element $g_e \in SU(2)$ to each edge in ${\cal
J}_{\va \Delta}$.  The action of the simplicial theory is given by
%\begin{equation}\label{action}
$S=\sum_{f\in {\cal J}_{\Delta}} B_{\ell_f} U_f$,
%\end{equation}
where $B_{\ell_f}$ is the Lie algebra element associated to the
$1$-simplex $\ell_f \in \Delta$, dual to the face $f\in {\cal
J}_{\Delta}$, and $U_f=g^1_eg_e^2\cdots g_e^N$ is the discrete
holonomy around $f\in {\cal J}_{\Delta}$. Since $\ell_f$ is in
one-to-one correspondence with $f\in {\cal J}_{\Delta}$, from now on
we denote $B_{\ell_f}$ simply as $B_f$.

The partition function is defined as
\begin{equation}\label{Zdiscrete}
{\cal Z}(\Delta)=\int \prod_{f \in {\cal J}_{\va \Delta}} \d B_f \
\prod_{e \in {\cal J}_{\va \Delta}} \d g_e  \exp(\i \Tr\left[B_f U_f\right]),
\end{equation}
where now $\d B_f$ is the regular Lebesgue measure on $su(2)\cong\R^3$,
$\d g_e$ corresponds to the invariant measure on $SU(2)$.

Integrating over $B_f$, we obtain
\begin{equation}\label{Zdiscrete0}
{\cal Z}(\Delta)=\int \ \prod_{e \in {\cal J}_{\va \Delta}} \d g_e \
\prod_{f \in {\cal J}_{\va \Delta}}{\huge \delta}(g^1_e\dots
g^{\va N}_e),
\end{equation}
where $\delta$ corresponds to the delta distribution defined on
${\cal L}^2(SU(2))$.

The integration over the discrete connection ($\prod_e \d g_e$) can be
performed if one expands the delta function in the previous equation
using harmonic analysis on the group.
% In the case of compact groups
%this is known as Peter--Weyl theorem, which asserts that any function
%$f\in {\cal L}^2(SU(2))$ can be written as a sum over matrix elements
%of unitary irreducible representations of $SU(2)$. 
Using the
Peter--Weyl decomposition, the $\delta$-distribution becomes
\begin{equation}\label{deltarep}
\delta(g)=\sum \limits_{j \in {\rm irrep}(SU(2))} \Delta_{j} \
\Tr\left[ \rho_j(g)\right],
\end{equation}
where $\Delta_{j}$ denotes the dimension of the unitary
representation $j$, and $\rho_j(g)$ is the corresponding representation
matrix. 
%Using equation (\ref{deltarep}), 
The partition function
(\ref{Zdiscrete0}) becomes
\begin{equation}\label{coloring}
{\cal Z}(\Delta)=\sum \limits_{{\cal C}_f:\{f\} \rightarrow \{ j\}}
\int \ \prod_{e \in {\cal J}_{\va \Delta}} \d g_e \ \prod_{f \in
{\cal J}_{\va \Delta}} \Delta_{j_f} \ \Tr\left[\rho_{j_f}(g^1_e\dots
g^{\va N}_e)\right],
\end{equation}
where ${\cal C}_f\colon\{f\} \rightarrow \{ j\}$ represents the
assignment of irreducible representations to faces in the dual
2-complex ${\cal J}_{\va \Delta}$. Each particular assignment is
referred to as a {\em coloring}. The summation is then over
colored $2$-complexes (spin foams).

If the $SU(2)$ group element $g_e$ corresponds to an $n$-valent
edge $e\in {\cal J}_{\va \Delta}$, i.e., an edge bounding $n$
faces, there are $n$ representation matrices evaluated on $g_e$ in
(\ref{Zdiscrete0}). The relevant integral is
\begin{equation}\label{3dp}
\int dg\ {\rho_{j_1}(g)}\otimes \rho_{j_2}(g) \otimes \cdots \otimes
\rho_{j_n}(g)= \sum_{\iota} {C^{\va \iota}_{\va j_1 j_2 \cdots j_n} \
C^{*{\va \iota}}_{\va j_1 j_2 \cdots j_n}},
\end{equation}
i.e., the projector onto ${\rm Inv}[\rho_{j_1}\otimes \rho_{j_2}
\otimes \cdots \otimes \rho_{j_n}]$. On the RHS we have chosen an
orthonormal basis of invariant vectors (intertwiners) to express the
projector.  Integrating over the connection (\ref{Zdiscrete0}) becomes
\begin{equation}\label{statesum}
{\cal Z}(\Delta)=\sum \limits_{ {\cal C}_f:\{f\} \rightarrow \{ j\}
} \ \sum \limits_{ {\cal C}_e:\{e\} \rightarrow \{ \iota\} }\
\prod_{f \in {\cal J}_{\va \Delta}} \Delta_{j_f} \prod_{v\in {\cal
J}_{\va \Delta}} A_v(\iota_v,j_v),
\end{equation}
where $A_v(\iota_v,j_v)$ is given by the appropriate trace of the
intertwiners $\iota_v$ corresponding to the edges bounded by the
vertex and $j_v$ are the corresponding representations. This
amplitude is given in terms of $SU(2)$ $3Nj$-symbols. When
$\Delta$  is a simplicial complex then all the edges in ${\cal
J}_{\Delta}$ are $3$-valent and vertices are $4$-valent. Consequently,
there are 3 representation matrices for all edges in (\ref{3dp})
and the corresponding amplitude is given by the contraction of the
corresponding four $3$-valent intertwiners, i.e., a $6j$-symbol.
In that case the partition function takes the familiar
Ponzano--Regge form
\begin{equation}\label{statesum1}
{\cal Z}(\Delta)=\sum \limits_{ {\cal C}_f:\{f\} \rightarrow \{ j\}} \
\prod_{f \in {\cal J}_{\va \Delta}} \Delta_{j_f} \prod_{v\in {\cal
J}_{\va \Delta}} \begin{array}{c}
\includegraphics[width=2.5cm]{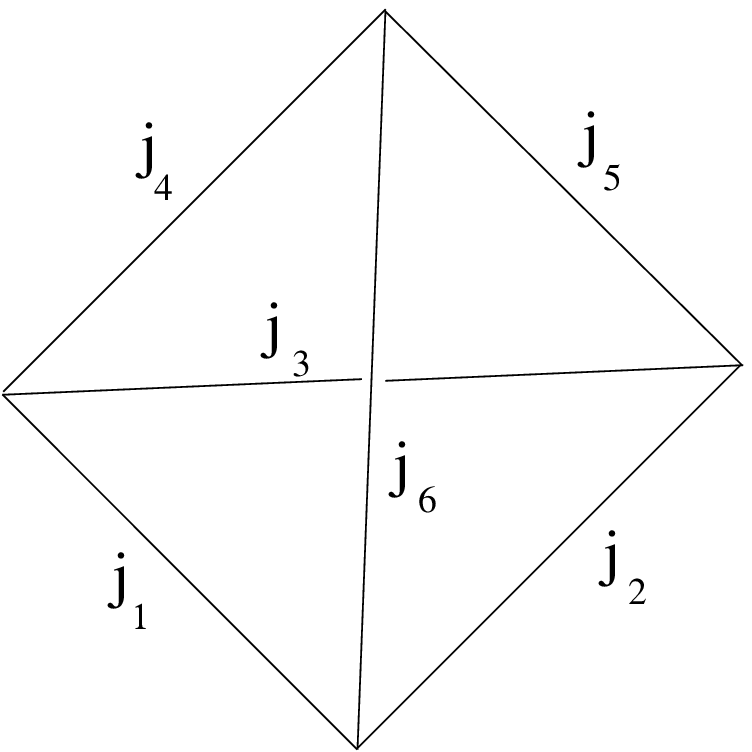}\end{array},
\end{equation}
where the tetrahedron corresponds to the graphical representation of
the $6j$-symbol.

In the next section we will analyze the case of gravity in four
dimensions.  The models of interest are defined in terms of
constrained 4-dimensional $BF$-theory which for Euclidean signature
has the gauge group $SO(4)\cong SU(2)\times SU(2)$. The discretization
of $BF$-theory in four dimensions is analogous to that of
3-dimensional $BF$-theory. The main difference is that the $B$ field
is now a Lie algebra valued $2$-form and so is discretized by the
assignments of Lie algebra elements $\{B_{t_f}\}$ to the 2-dimensional
surfaces defined by the triangles $t_f\in \Delta$.  Triangles are in
one-to-one correspondence with faces $f\in {\cal J}_{\Delta}$. The
connection is discretized in precisely the same way as in three
dimensions; namely, by assigning group elements $g_e$ to edges $e\in
{\cal J}_{\Delta}$. Upon integration over $\{B_{f}\}$ and the
$\{g_e\}$ the amplitudes can be expressed as a spin foam sum similar
to (\ref{statesum1}), where the $6j$-symbol is replaced by a
$15j$-symbol represented by a 4-simplex (see \cite{review} for details
and references). Since any $SO(4)$-representation can be decomposed
into a product of two $SU(2)$-representations, the face amplitude is now
$\Delta_{(j_1,j_2)}=(2j_1+1)(2j_2+1)$.

\subsection{Gauge fixing for 3d gravity}

The expression (\ref{statesum}) is generically divergent due to
%.  A reason for this divergence is 
the (non-compact) gauge freedom (\ref{gauge2})
\cite{fre7}.  Some of the $B$ integrations in
(\ref{Zdiscrete}) -- or equivalently some of the $\delta$-functions in
(\ref{Zdiscrete0}) -- are thus redundant.
% The explicit form of the discrete
%gauge symmetry of simplicial 3-dimensional gravity is described in
%detail in \cite{fre7}.
In addition to the standard $SU(2)$ gauge invariance of the action
corresponding to (\ref{gauge1}), there is a discrete analog of
(\ref{gauge2}) given by
%Namely, the simplicial action is invariant under the transformation
\begin{eqnarray}\label{topo}
 \delta B_{\ell_f}=\eta_v -[\Omega_{\ell_f}^v, \eta_v ]
 \mbox{ if } v\subset \ell_f \qquad \mbox{ and }\qquad \delta
 B_{\ell_f}=0 \mbox{ if } v \not\subset \ell_f
\end{eqnarray}
where $\eta_v$ is a Lie algebra element associated to the vertex $v\in
\Delta$ and $\Omega_{\ell_f}^v$ is also in the Lie algebra and can be
explicitly given in terms of the logarithm of the elements
$\{U_{f^{\prime}}\}$ for $f^{\prime}\not= f$ and contained in the set
of faces that form the dual bubble in ${\cal J}_{\Delta}$ around the
vertex $v$ \cite{fre7}.  The previous transformation is a symmetry of
the simplicial action due to the discrete version of the Bianchi
identity stating that the (ordered) product of $U_f$ around a bubble
is equal to the identity.

Assuming $\Delta$ is path connected we can set $B=0$ along a
contractible (within $\Delta$) path $L$ containing all vertices in
$\Delta$ using the gauge freedom (\ref{topo}).  This fixes this gauge
freedom completely
%.  As shown in \cite{fre7} this gauge fixing
and contributes to the measure with the Faddeev--Popov determinant
% of the type appearing in (\ref{Zsec1})
\begin{equation}\label{fp}
\det(\partial f_I/\partial\delta_J|_{\delta=0})=(1+|\Omega^{v}_{\ell_f}|^2)
\end{equation}
which only depends on the connection $\{g_e\}$. Integration over the
$B$ field produces the curvature delta functions as in
(\ref{Zdiscrete0}) which in turn imply $\Omega^{v}_{\ell_f}=0$ and
hence trivial Faddeev--Popov factors.

The result is very simple: we have to drop out all the
$\delta$-functions in (\ref{Zdiscrete0}) corresponding to faces $f \in
{\cal J}_{\Delta}$ dual to $1$-simplexes in $L$ which are
redundant leading to divergences in the non gauge fixed
formulation. The gauge fixing is then analogous to changing the
discretization. In this precise sense we find a connection between
discretization independence of the partition function and the gauge
freedom (\ref{topo}).

\section{Spin foam measure}\label{GR}

As we have seen in the previous section, the well-known amplitudes of
the spin foam model of $BF$-theory can be seen as emerging from a
concrete transition from the continuum measure of a path integral. In
gravity models which are defined as constrained $BF$-theories (of the
type of Barrett and Crane's), this direct construction of the measure
is not available.  This is so because the simplicity constraints are
imposed on $BF$ amplitudes after the integration over the (discrete)
$B$-field has been performed. At this stage, one is already dealing
with a discrete state sum--where $B$ configurations are replaced by
irreducible unitary representations--and the connection with the
formal continuous measure is lost.

In this section, we will first present a method to impose the
simplicity constraints (or, more generally, second class
constraints in a spin foam quantization) in such a way that there
is always a clear connection to the continuum measure. It is based
on a passive interpretation where the (diffeomorphism) gauge
transformations do not change the discretization, but rather
values of the fields.  Explicit calculations in the case of
gravity, however, so far look complicated and we will not pursue a
calculation of the amplitudes here \cite{Buffenoir:2004vx}.
Nevertheless, one can hope that at least their asymptotic behavior
for large labels can be found easily, which would allow us to see
whether or not the state sum will be finite.

Independently, we can use the active picture where diffeomorphisms
change the discretization. There are now moves which lead to a
different discretization but a configuration which has to be
considered as physically equivalent to the original one.
Background independence then requires that the amplitudes do not
change under those moves, which as we will see in
Subsection~\ref{DC} imposes non trivial restrictions on the
measure. In Subsection~\ref{te} we present a toy model of a
background independent spin foam.

Anomaly freedom and background independence both restrict the lower
dimensional amplitudes of a spin foam. Though not necessarily
equivalent, both notions are clearly related since a model which has
an anomaly for the diffeomorphism gauge group cannot be background
independent, and vice versa. We expect the conditions for
the amplitudes to have similar consequences, which can serve as a
non-trivial consistency check for the calculations. This will be shown
later to be realized in the Husain--Kucha\v{r} model.

\subsection{Passive picture}

%As recalled in Section~\ref{qbf}, 
In the standard derivation of the
spin foam models for $BF$-theory there is a clear-cut relationship
between the path integral measure and spin foam amplitudes
%. This relationship 
coming directly from integrating the
$B$-field in (\ref{Zdiscrete})
% which results in (\ref{Zdiscrete0}) 
and
using the Peter--Weyl decomposition of the $\delta$-distribution.
% on the group (harmonic analysis on the group).  
%We can relate the $B$
%configurations with the spin labels appearing in the state sum in a
%way that can be useful for exploring the definition of the spin foam
%measure in the more complicated case of gravity. 
In the next subsection we will re-derive part of the spin foam
quantization of $BF$-theory with different methods which show how the
labels $j$ arise from a discretization of $B$. This will allow us to
propose a recipe for the construction of the correct measure in the
case of gravity in Subsection~\ref{GRe}.
%The implementation of this general
%construction requires the knowledge of the constraint structure of the
%discrete theory that is not available at present.

\subsubsection{$BF$-theory in polar coordinates}
\label{BFp}

%A spin foam quantization of gravity is supposed to include, before
%taking the continuum limit, the main steps already encountered in our
%toy model: replacing the integral representation of $\delta$-functions
%with a series representation (which discretizes the $B$-field) and
%solving the simplicity constraints by restricting the summation
%labels. However, the actual derivation proceeds along different lines
%which simplifies the calculations drastically but also obscures the
%relation between the original path integral measure and the resulting
%face amplitude. Since $BF$-theory is an important intermediate result,
%we will first discuss its spin foam quantization and show that its
%usual face amplitude is a direct consequence of the invariant path
%integral measure.

%As recalled in Section \ref{qbf}, 
A main ingredient of the spin foam
quantization of $BF$-theory in three or four dimensions is the formula
\begin{equation}\label{delta}
 (2\pi)^{-1}\int\d B\exp(2\i\tr(Bg))=\delta(g)=\sum_j(2j+1){\rm
 tr}\rho_j(g)
\end{equation}
where $g\in SU(2)$, $B\in su(2)$.
%, which follows from the Peter--Weyl theorem. 
%By using this formula, 
The continuous values of $B$ are effectively replaced by discrete
values $j$, but the exact correspondence remains unclear. In
particular, there are three independent components in $B$, but only
one discrete label $j$. The following calculations will now show that
$j$ is the discretized radial component of $B$ in polar coordinates
whereas the angular components are integrated out.

We write the $su(2)$ element $B=rn^i\tau_i$ with $n=
(\sin\vt\cos\vp,\sin\vt\sin\vp, \cos\vt)$ in polar coordinates
$(r,\vt,\vp)$ where $\tau_j=\i /2\sigma_j$ with Pauli matrices
$\sigma_j$. To simplify the calculation we also choose the $SU(2)$
element in the gauge $g=\exp
c\tau_3=\cos\frac{c}{2}+2\tau_3\sin\frac{c}{2}$ without loss of
generality (thanks to gauge invariance of the trace in
(\ref{delta})). This leads to
\begin{eqnarray} \label{deltacont}
 \fl\delta(g) = (2\pi)^{-1}\int\d r\d\vt\d\vp r^2\sin\vt\e^{-2\i r\cos\vt
 \sin c/2}
%\nonumber\\
 = \i\int\d r\,\frac{r}{2\sin\case{c}{2}}
 (\e^{-2\i r\sin c/2}- \e^{2\i r\sin c/2})
\end{eqnarray}
%\begin{eqnarray} \label{deltacont}
% \fl\delta(g) = (2\pi)^{-1}\int\d r\d\vt\d\vp r^2\sin\vt\exp(-2\i r\cos\vt
% \sin\case{c}{2})\nonumber\\
% \lo= \i\int\d r\,\frac{r}{2\sin\case{c}{2}}
% (\exp(-2\i r\sin\case{c}{2})- \exp(2\i r\sin\case{c}{2}))
%\end{eqnarray}
whereas the discrete form is
\[
 \fl\delta(g)=\sum_j(2j+1)\frac{\sin(j+\case{1}{2})c}{\sin\case{c}{2}}=
 \sum_j\frac{2j+1}{2i\sin\case{c}{2}}(\exp(\i (2j+1)\case{c}{2})-
 \exp(-\i (2j+1)\case{c}{2}))\,.
\]
%A comparison shows that 
One thus obtains the discrete version of the $\delta$-function by
integrating out $\vt$ and $\vp$, and replacing the continuous $r$ with
the discrete $j+\frac{1}{2}$. (One also has to replace
$\sin\frac{c}{2}$ with $\frac{c}{2}$, but since both expressions
represent a $\delta$-function in $c$, they can be regarded as
identical.) This clarifies the relation between continuous values for
$B$ and the discrete $j$.

The calculation can also be used to show that the standard $BF$-face
amplitude agrees with the one derived from the invariant path integral
measure and thus is anomaly-free. For Euclidean gravity, $SO(4)$
$BF$-theory is used which can be written as a state sum with
$SU(2)$-valued variables after using $SO(4)\cong SU(2)\times
SU(2)$. For each copy of $SU(2)$ on every face with spin $j_f$ one has
a factor contributing to the face amplitude by $2j_f+1$ as
in the last result in (\ref{deltacont}) where a single factor $r$
remains in the measure after integrating over $\vt$. Discretizing $r$
then yields the correct face amplitude (Section \ref{qbf}).

\subsubsection{A general recipe} \label{GRe}

For theories more complicated than $BF$-theory, but still written as
constrained $BF$-theory, we have to combine the result of
Section~\ref{measure}, which tells us the correct continuum measure in
the presence of second class constraints, with a transition from the
continuum measure to discrete amplitudes. Since the analysis must be
at the canonical level and some field components are distinguished as
Lagrange multipliers, in general such a measure will break manifest
covariance. We also have to expect additional functions which result
from inserting $\delta$-functions imposing the constraints into the
path integral. The constraint algebra tells us what function we have
to include to obtain the correct continuum measure as in
Eq.~(\ref{Zsec}), and there can be additional functions coming from
Jacobians if the constraints do not directly restrict the integration
variables but a more complicated functional of them.  The constraint
structure of Plebanski's theory has been studied in detail in
\cite{Buffenoir:2004vx}.

A possible strategy is to use the formulation of $BF$-theory in polar
$B$-coordinates where we have seen that the spin foam model with the
correct face amplitude arises from integrating out the $B$-angles and
discretizing the radial $B$-coordinate. Additional constraints
restricting the $BF$-theory then arise from the action via integrating
over the corresponding Lagrange multipliers which results in
$\delta$-functions inserted into the path integral. We have seen that the
multiplier integration requires a special measure which can be
computed if the constraint algebra is known. Next, we write both this
function and the $\delta$-functions in the polar $B$-coordinates,
integrate over the $B$-angles and discretize the radial $B$-coordinate
just as we did in $BF$-theory without any additional constraints. In
general, this will result in an additional functional besides the
vertex amplitude in the spin foam model which depends on the spins $j$.

In practice, the calculation, in particular the computation of the
constraint algebra and the integration over the angular
$B$-coordinates, will be complicated (several faces are coupled by the
simplicity constraint). To illustrate the type of integrals involved
we have included an example of constrained $BF$-theory in the
appendix.  It is also not easy to decide which part of the constraint
algebra one has to consider. All constraints taken together form a
mixed system containing first class constraints. Usually, as in
Section~\ref{measure}, one would have to gauge fix the first class
constraints resulting in a pure second class algebra. However, for
gravity a gauge fixing is not known, and it is hoped that spin foam
models can avoid the need to introduce an explicit gauge fixing by
using a space-time discretization. In the following subsection we will
see however that simple considerations of background independence with
respect to active diffeomorphisms imply the existence of bubble
divergences which are naturally associated to infinite gauge volume
contributions.  Diffeomorphisms in fact seem not to be completely
gauge fixed by the discretization and the situation might be similar
to that of $BF$-theory reviewed in the previous sections.  One then
would have to find a second class sub-algebra of the full constraint
system plus the appropriate gauge fixing constraint in order to
compute the correct measure.

It seems necessary to look for a description of gauge symmetries and
constraint algebra that would be directly defined at the discrete
level.  In the previous section we have reviewed how this can be done
in the case of $BF$-theory. Indeed, in order to regularize the path
integral Freidel and Louapre had to introduce a gauge fixing of the
topological symmetry (\ref{topo}) which in turn modifies the measure
by the Faddeev-Popov determinant (\ref{fp}).  In the case of this
topological theory the modification is trivial and the factor is
independent of the fields reducing simply to unity.  In the case of
gravity one should expect a non trivial dependence on spins consistent
with a theory with local excitations.  It is clear from this example
that even when the gauge symmetries of the discrete action retain some
similarities with the continuum ones their action can be only
interpreted at the discrete level. An equivalent analysis in the case
of 4-dimensional Plebanski theory seems necessary in order to
implement the general prescription of Subsection~\ref{GRe} for the
construction of the measure and hence settle the issue of lower
dimensional simplex amplitudes.  In such a context, the first class
part of the constraint algebra might not even be first class in the
continuum sense as results of Gambini and Pullin show
\cite{pul0,pul00}.  In this case a direct application of our recipe
(or a slight generalization) should be feasible.

%Even if not all calculations can be done explicitly at this stage, one
%can still try to extract information as, e.g., the asymptotic behavior
%of the face amplitude as a function of $j$. This would be necessary to
%decide whether or not the model is finite on a fixed triangulation ---
%even without a gauge fixing. As an example we consider a degenerate
%model of gravity where some of the calculations are simpler, and end
%with a few remarks about full gravity.
%end of new part

\subsection{Active picture}

%Our aim is to derive conditions for the correct path integral measure
%for gravity in the context of spin foam models. 
In this section we
will show how the requirement of background independence and
diffeomorphism invariance from the active point of view imposes
restrictions on the spin foam amplitudes with important
consequences. 
%The discussion presented here is general to any spin
%foam model of a diffeomorphism invariant theory. However,
For concreteness, we will focus the attention on the Barrett--Crane
model for quantum gravity
%. We shall perform our derivation in the
%context of 
in the formulation of simplicial quantum gravity presented in
\cite{a16}.
% In this formulation the Barrett--Crane model arises as the
%simplicial counterpart of Plebanski's formulation of gravity.

\subsubsection{Background independence}\label{DC}

\begin{figure}[h]
\centerline{\hspace{0.5cm} \(\begin{array}{c}
\includegraphics[width=3.8cm]{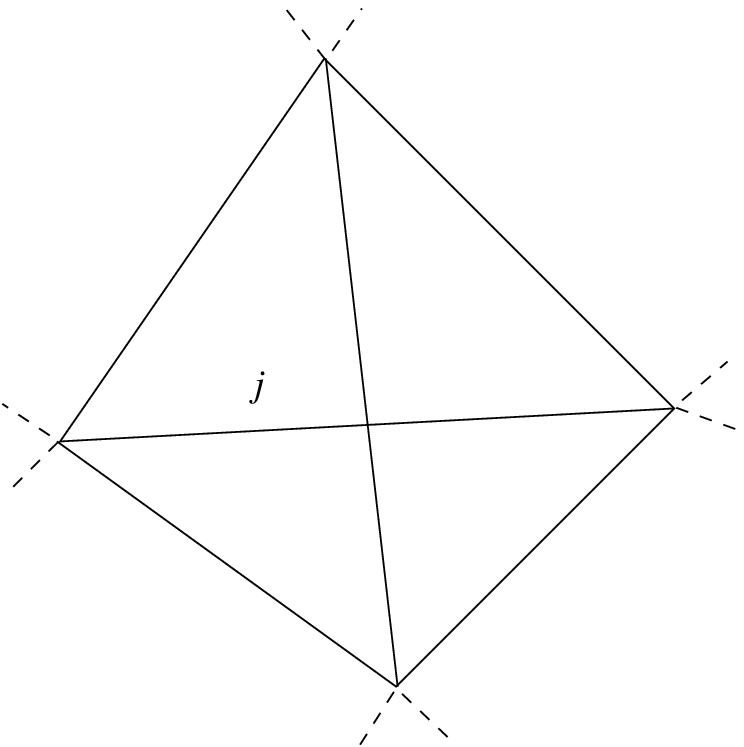}\end{array}\begin{array}{c}
\begin{array}{c}
\rightarrow \\ {\rm (ii)}\end{array}
\end{array}
\begin{array}{c}\includegraphics[width=3.8cm]{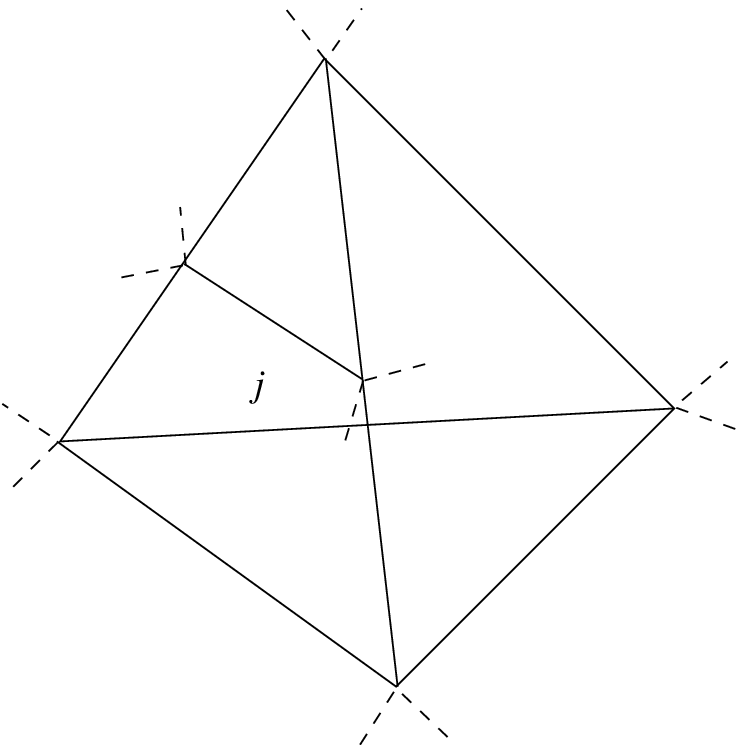}\end{array}
\begin{array}{c}
\end{array}  \begin{array}{c}
\rightarrow \\ {\rm (i)}\end{array}\begin{array}{c}
        \includegraphics[width=3.8cm]{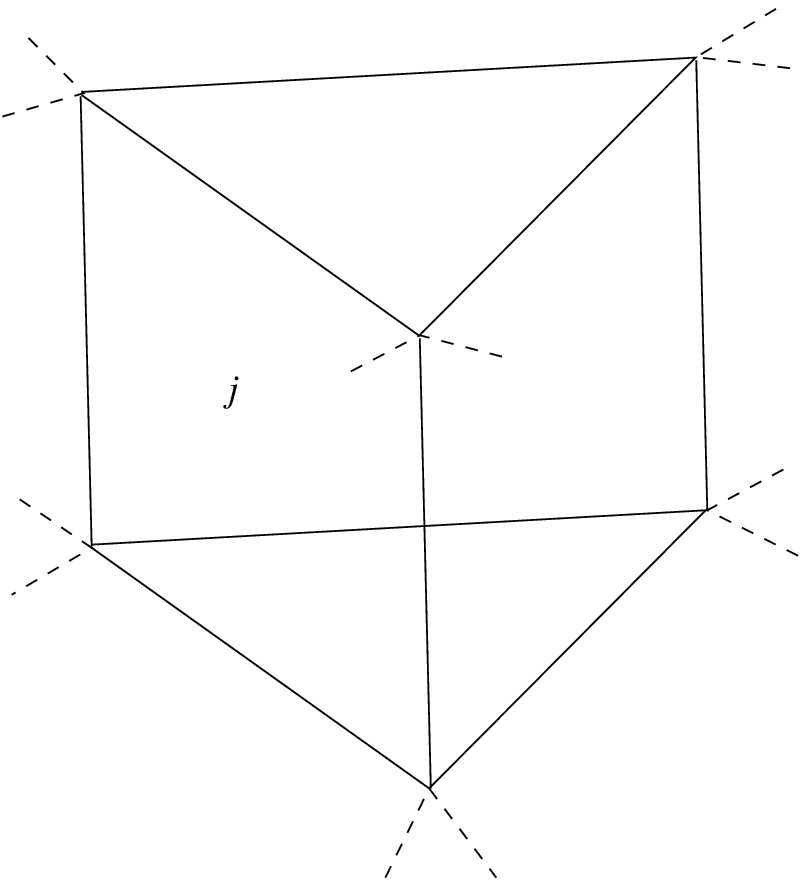}\end{array} \) }
\caption{Vacuum bubbles to be physically equivalent in an anomaly
free spin foam model. Their equivalence constrains the possible
behavior of the face and edge amplitudes. We represent the sequence of moves
that relate the bubble on the left with that on the right.} \label{vb}
\end{figure}

The requirements imposed here on the spin foam amplitudes can be
viewed as a 4-dimensional generalization of the notion of
cylindrical consistency and diffeomorphism invariance in the
canonical formulation of loop quantum gravity \cite{th4}. According to
background independence, the cellular decomposition used to
represent the space-time manifold does not carry any physical
information. Gravitational degrees of freedom are encoded in the
labeling of faces in the dual 2-complex with irreducible
representations of the corresponding internal gauge group: a spin
foam. As a consequence there remains some redundant information in
a spin foam defined on a particular 2-complex which links the
`physical' configuration with the discretization on which it has
been defined. The background independent information is encoded in
appropriate equivalence classes of spin foams. These
equivalence classes have to be introduced  if one wants to think
of spin foams as morphisms in the spin network category
\cite{baez7}, and are defined
%. Elements of a given equivalence class of spin foams
%can be related 
by the following moves:
\begin{enumerate}
\item {\em (Piecewise linear) maps preserving the cell-complex
and its coloring}
\item {\em Subdivision}
\item {\em Orientation reversal}
\end{enumerate}
A detailed definition can be found in \cite{baez7}. These moves
can be interpreted as the counterpart of diffeomorphisms in
simplicial quantum gravity (with perhaps the addition of more
equivalence relations if the remnant of diffeomorphism invariance
is larger as in $BF$-theory). Transformations given by arbitrary
combinations of the moves are clearly limited and the resulting
equivalence relation is much weaker than complete discretization
independence which would be the requirement for a topological
quantum field theory. The most important move is the second one
since the others leave the number and connections of simplices
invariant. An illustration of the limited scope of the moves can
be seen in Fig.~\ref{vb} where only one new edge and two new
vertices are created while the rest is unchanged. Thus, the
discretizations have to be regarded as defining equivalent
backgrounds which are just parameterized differently. (This point
is different from lattice gauge theories in a given background
where a new edge would probe new degrees of freedom.)

%The situation is analogous to that of the canonical formulation of
%loop quantum gravity where one essentially solves the diffeomorphism
%constraint by considering equivalence classes of spin networks under
%3-diffeomorphisms. The spin foam equivalence classes correspond to the
%4-dimensional generalization of this idea. In the following we will
%see how these considerations restrict to some extent the freedom in
%the definition of the spin foam measure.

The natural amplitude for faces in spin foam models is given by
the Plancherel measure arising in the harmonic analysis on the
corresponding internal gauge group.
% (e.g., $Spin(4)$ and $SL(2,C)$
%for the Riemannian and Lorentzian models respectively). 
This can
be derived in various ways and it is directly linked with the
notion of locality of spin foams \cite{reis4}, namely that degrees
of freedom communicate along faces by boundary data given by the
%$Spin(4)$ and $SL(2,C)$ 
connection.
% respectively. 
If the face
amplitude $A_f(j)$ is given by the Plancherel measure (e.g.,
$A_f(j)=(2j+1)^2$ in the case of the Riemannian Barrett--Crane
model) then the arbitrary subdivision of a face $f\in {\cal
J}_{\Delta}$ does not change the amplitude. \footnote{In the
canonical picture of loop quantum gravity the boundary data is
given by $SU(2)$ connections for which the Plancherel measure
iduces a face amplitude $A_f=2j+1$. In spin foam models based on
$Spin(4)$ $BF$-theory such a face amplitude could arise as a
result of the implementation of second class constraints as
discussed in Section \ref{GRe}.} 
%Amplitudes are invariant under
%subdivision of faces which produces combinatorially equivalent
%spin foams. Assigning different amplitudes to these would
%correspond to an anomaly (in the sense that members of the same
%equivalence class would be associated with different amplitudes).
%Recall that in background independent spin foam models the
%2-complex has no geometrical meaning whatsoever and two spin foams
%for which a face has been subdivided in this way correspond to
%physically equivalent configurations.

\begin{figure}[h]
\centerline{\hspace{0.5cm} \(\begin{array}{c}
\includegraphics[width=4cm]{tet1.eps}\end{array}
\begin{array}{c}
\end{array} \begin{array}{c}
\rightarrow \\ {\rm (ii)}\end{array}
\begin{array}{c}\includegraphics[width=4cm]{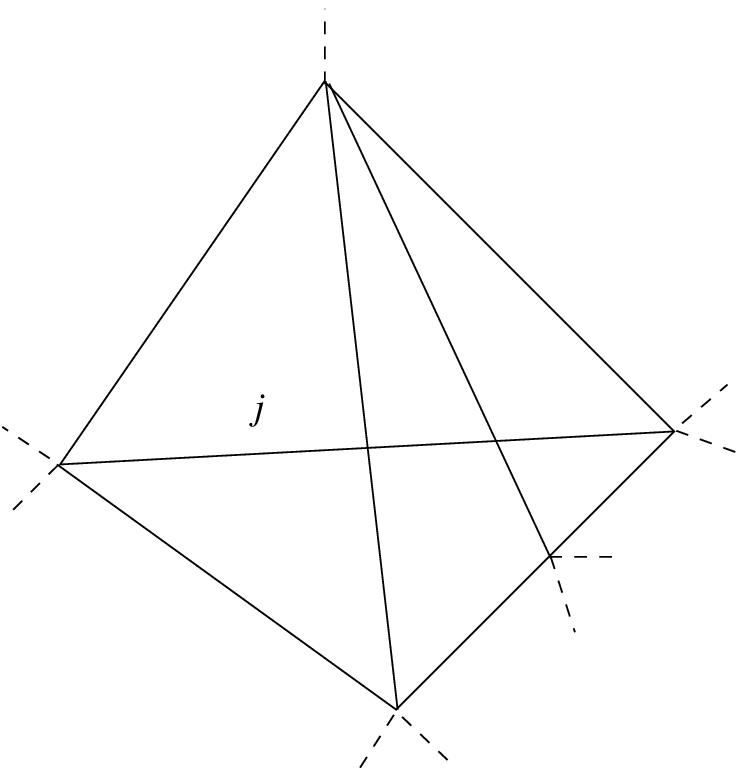}\end{array}
\begin{array}{c}
 \end{array}   \begin{array}{c}
\rightarrow \\ {\rm (i)}\end{array}\begin{array}{c}
        \includegraphics[width=4cm]{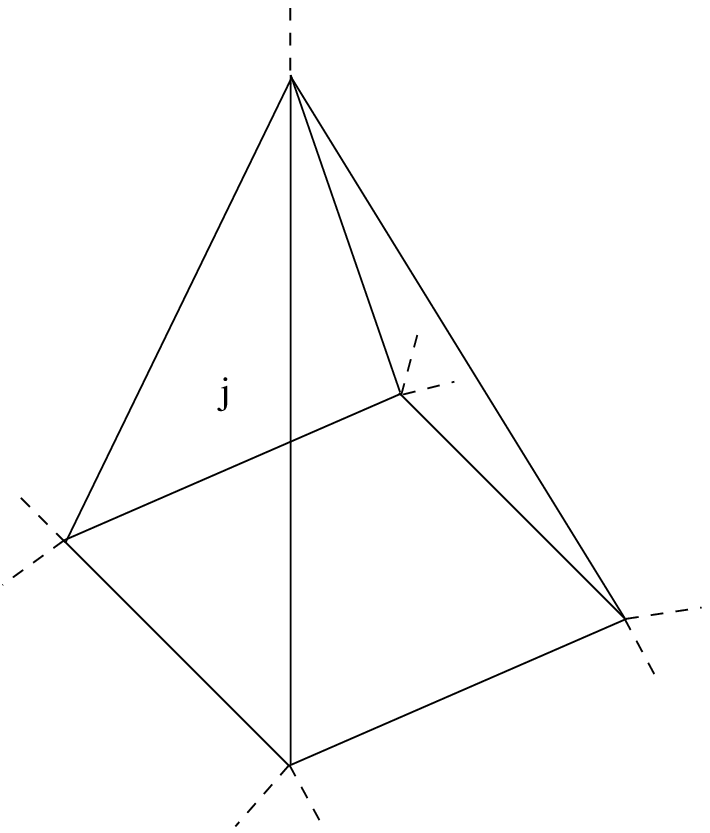}\end{array} \) }
\caption{Sequence of subdivision and piecewise linear transformation
relating the tetrahedral bubble with the pyramidal one.} \label{vb1}
\end{figure}

The previous analysis raises the question of whether we can find
more stringent conditions on spin foam amplitudes by solely
imposing background independence.
% In fact this is possible. We can
%obtain restrictions on both the face and edge amplitudes by
%imposing further necessary conditions for the background
%independence of the spin foam model.  In fact, we have seen that
%the non trivial action of gauge transformation in $BF$-theory
%suggests that even in a simplicial theory a non trivial remnant of
%the diffeomorphisms might survive. The precise nature of this
%remnant is not understood but there are some basic symmetries that
%have to be represented there. These correspond to the discrete
%symmetries of the spin foams which are imposed by the requirement
%of background independence. As discussed above, two spin foams are
%to be thought of as physically equivalent whenever we can `deform'
%one into the other by the action of one of the moves mentioned at
%the begining of this section \cite{baez7}.
% A necessary consistency
%condition is that amplitudes be well defined on the equivalence
%class of spin foams.
%
The simplest consistency condition can be obtained by the
requirement that the bubble spin foams of Figures \ref{vb} and
\ref{vb1} have the same amplitude. The figure represents spin
foams where most of the faces are labeled by the trivial
representation except for the shown bubbles, labeled by the simple
representation $\rho=j\otimes j^*$. These spin foams are clearly
equivalent under the above moves and the figures illustrate the
sequence of moves that relate them. This is also in agreement with
our intuitive notion of background independence (see \cite{review}
for more discussion): given that the underlying 2-complex does not
carry any geometrical information and that geometry degrees of
freedom are represented by the labeling of faces with spins and
its intrinsic combinatorics there is no background independent way
to distinguish between these bubble spin foam excitations. Their
apparent difference is linked to the fiducial background 2-complex
used to represent them and should not play any physical role.
(This is true only if, as assumed, all outside faces are labeled
by the trivial representation; otherwise they will not be related
by the moves defined above. Again, this illustrates that the
equivalence relation used here is much weaker than that for
topological field theories.)

%Let us compute these amplitudes in the case of the Barrett--Crane
%model.\footnote{The same analysis can be performed with any other
%model.} 
For example, the Barrett--Crane model is a definition of the
$4$-simplex or vertex amplitude for quantum gravity up to an
overall factor. If we normalize the vertex amplitude in some
arbitrary way we can shift the ambiguity to the value of the edge
amplitude. In addition we assume the edge amplitude to be local
in the sense that it only depends on the values of representations
that label the corresponding edge. This assumption is in
correspondence to the models defined in the
literature.\footnote{However, it is possible that non local
modifications of the amplitude could arise as a result of the
implementation of constraints as in Section \ref{GRe}. This
possibility will not be explored here.} The normalization we
choose is that for which the Barrett--Crane intertwiner is a
norm-one-vector in the Hilbert space where ${\rm Inv}[ \rho_1
\otimes\rho_2\otimes\rho_3\otimes\rho_4]$ acts. This normalization
is naturally obtained in the implementation of Plebanski's
constraints on the $BF$ partition function \cite{a16}.

\begin{figure}[h]
\centerline{\hspace{0.5cm} \(
\begin{array}{c}
        \includegraphics[width=4cm]{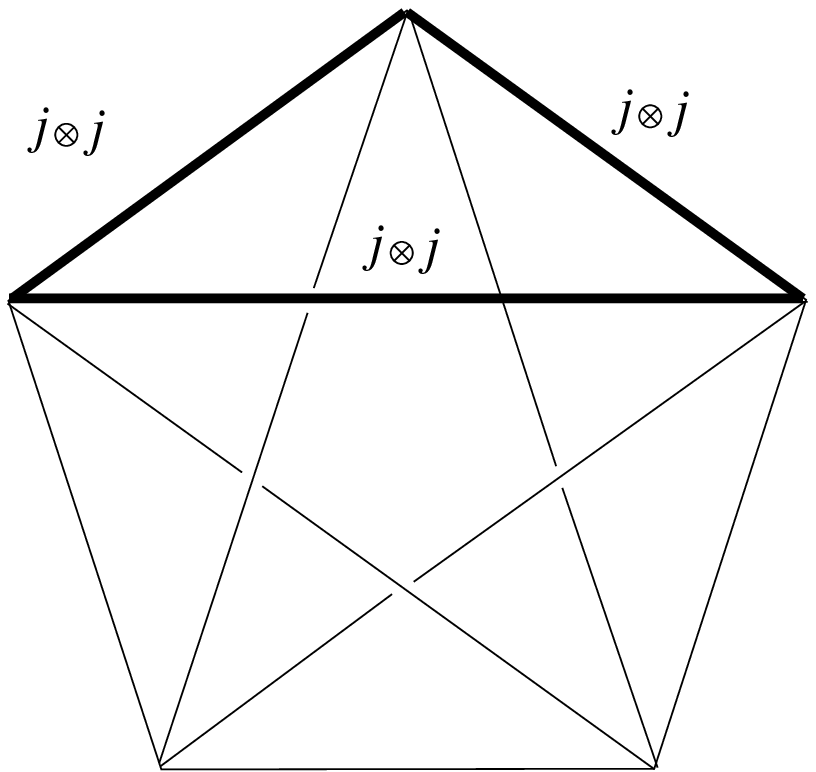}\end{array}
\ \ \ \ \ \ \ \ \ \ \begin{array}{c}
        \includegraphics[width=4cm]{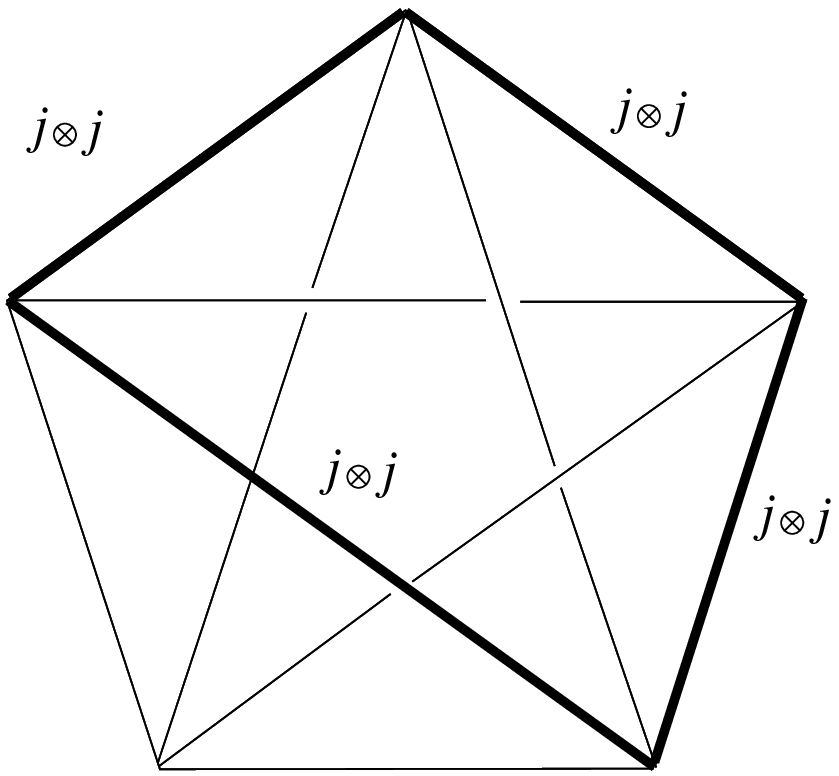}\end{array} \) }
\caption{Relevant 15j-symbols for $B_1(j)$, $B_2(j)$ and $B_3(j)$.
% in Equations (\ref{a}), (\ref{b}) and (\ref{c}). 
Thin lines are trivial representations.} \label{vb44}
\end{figure}

Let us first study the bubble amplitudes in Figure~\ref{vb}.  The
first bubble from left to right has four vertices corresponding to
the Barrett--Crane $10j$-symbol illustrated on the left diagram in
Figure \ref{vb44} containing only three edges labelled by non
trivial simple representations $j\otimes j$. The value of this
vertex can be easily evaluated \cite{review}. It corresponds to
the value of the trace of the three non-trivial normalized
intertwiners appearing in the triangular loop on the left of
Figure~\ref{vb44}.  If we denote by $\iota$ the corresponding
normalized 2-intertwiner it is obvious that
$\iota=(2j+1)^{-1} {\id}\otimes{\id}$, where ${\id}$ is the identity in
the vector space corresponding to the $SU(2)$ representation
$j$. Therefore the result is \[{\rm
Tr}[\iota\cdot\iota\cdot\iota]=\frac{1}{(2j+1)^3} {\rm Tr}[\id\otimes
\id]=\frac{1}{2j+1}.\] The amplitude of the tetrahedral bubble
is then $B_1(j)=(2j+1)^{-4}A_e(j)^6 A_f(j)^4$,
%\begin{equation}\label{a}
%B_1(j)=(2j+1)^{-4}A_e(j)^6 A_f(j)^4,
%\end{equation}
where $A_e(j)$ and $A_f(j)$ are the so far undetermined edge and face
amplitudes. The exponents are: $6$ for the six edges and $4$ for the
four faces of the tetrahedral bubble.  We have given already an
argument for the value of $A_f(j)$ which we will re-derive here from
the background independence condition. In the case of the bubble
diagram on the right of Figure \ref{vb} we have six vertices
corresponding to the same $10j$-symbol as before, so the amplitude for
the prism bubble is $B_2(j)=(2j+1)^{-6} A_e(j)^9 A_f(j)^5$.
%\begin{equation}\label{b}
%B_2(j)=(2j+1)^{-6} A_e(j)^9 A_f(j)^5.
%\end{equation}

Finally in the case of the bubble spin foam on the right of Figure
\ref{vb1} we have four vertices of the previous type plus the
vertex on the top whose $10j$-symbol is illustrated in the diagram on
the right of Figure \ref{vb44} which evaluates to $(2j+1)^{-2}$. With
this the amplitude of the pyramid bubble is $B_3(j)=(2j+1)^{-6}
A_e(j)^8 A_f(j)^5$.
%\begin{equation}\label{c}
%B_3(j)=(2j+1)^{-6} A_e(j)^8 A_f(j)^5.
%\end{equation}
The requirement $B_1(j)=B_2(j)=B_3(j)$ fixes the values of $A_e(j)$
and $A_f(j)$ uniquely to
\begin{equation}
A_e(j)=1 \ \ {\rm and} \ \ A_f(j)=(2j+1)^2. \label{amp}
\end{equation}

\begin{figure}[h]
\centering {\includegraphics[width=4cm]{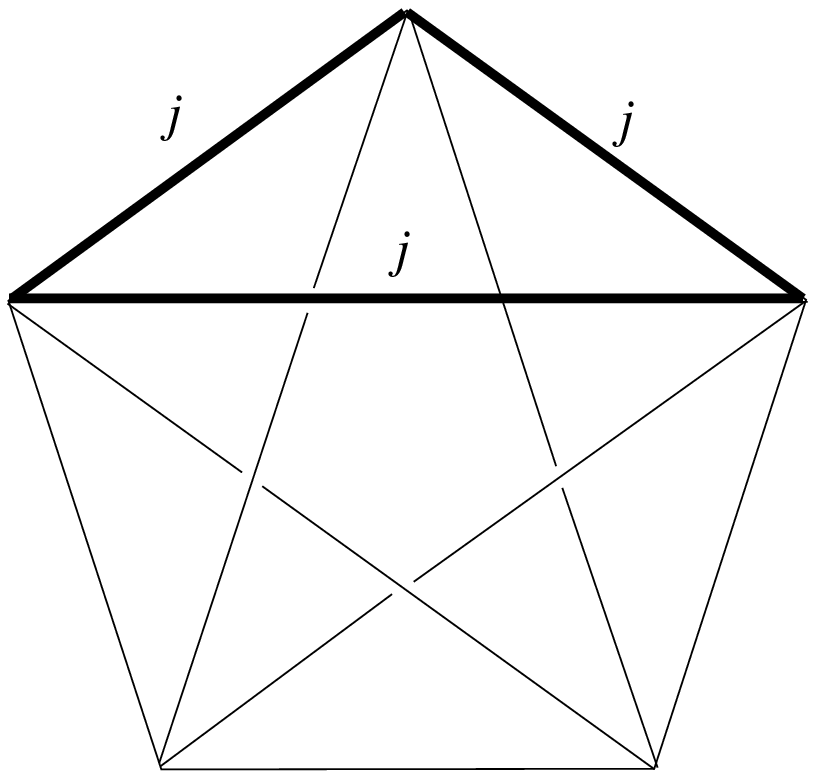}
            \includegraphics[width=4cm]{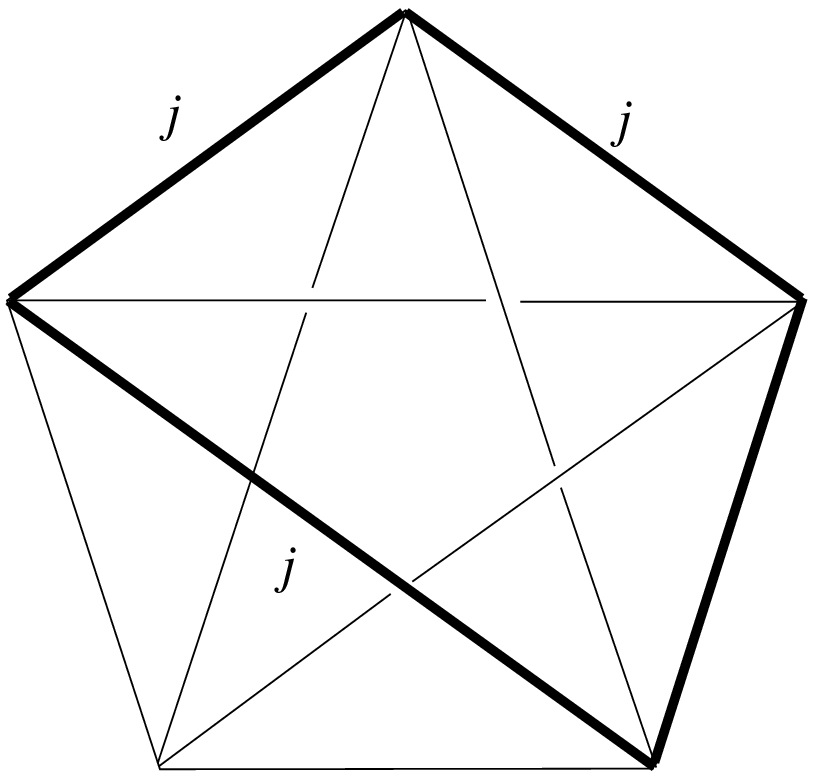}
        \includegraphics[width=4cm]{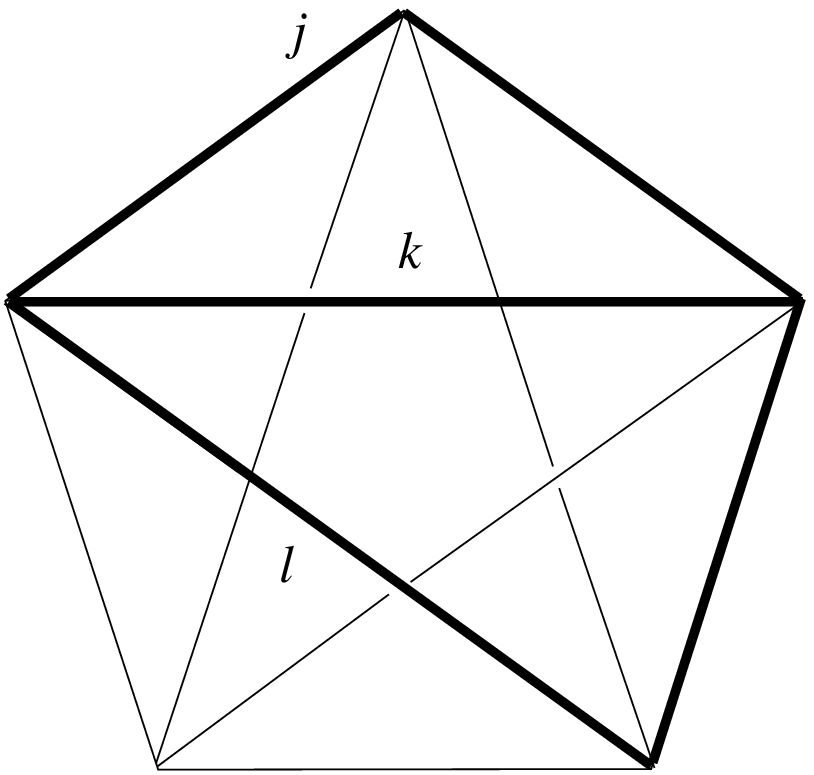}},
\caption{Vertex contributions to the bubble amplitudes above (thin
  lines represent edges labeled with the trivial representation). From
  left to right their value is given by $(2j+1)^{-1}$, $(2j+1)^{-2}$
  and $(2j+1)^{-1}(2l+1)^{-1}$ in the Riemannian Barrett--Crane model
  if we normalize the corresponding intertwiners.} \label{10j}
\end{figure}

\begin{figure}[h]\centerline{\hspace{0.5cm} \(
\begin{array}{c}\includegraphics[width=5cm]{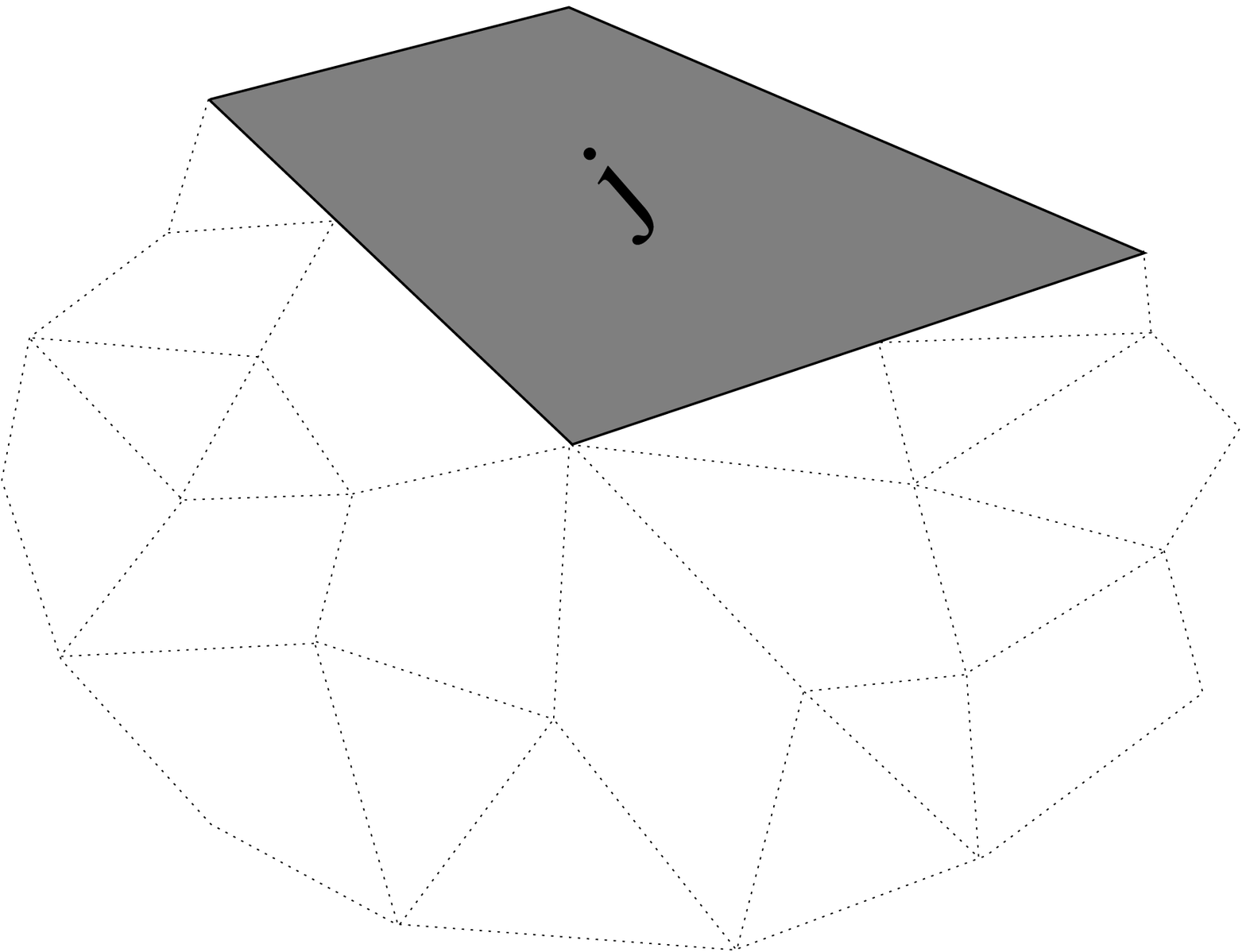}\end{array}
\begin{array}{c}
 \end{array}  \ \ \ \
\rightarrow \ \ \ \
\begin{array}{c}
        \includegraphics[width=5cm]{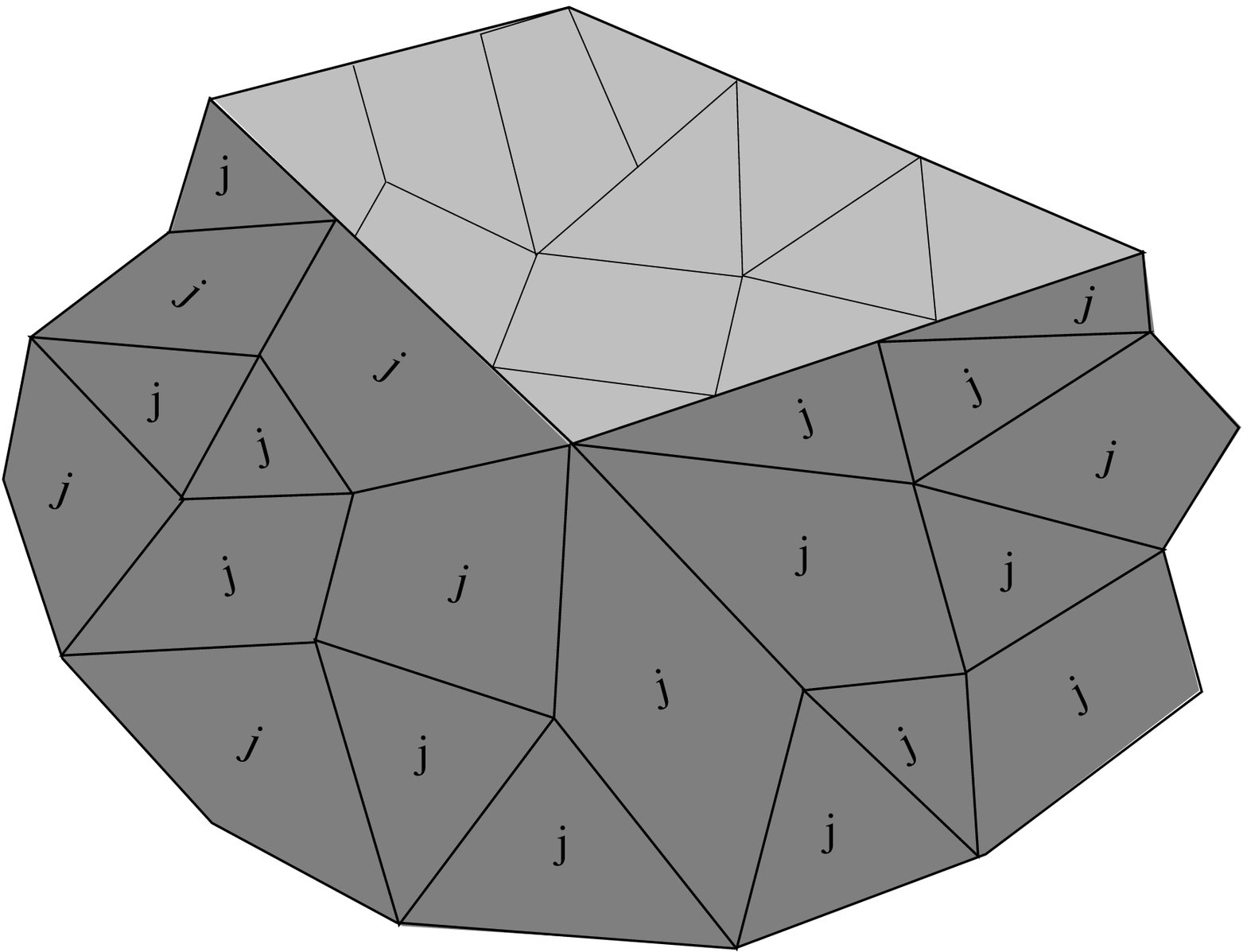}\end{array} \) }
\caption{Two equivalent spin foam configurations. The dotted 2-cells on 
the left are labelled by the trivial representation.
  } \label{subdi}
\end{figure}

The previous bubble spin foams are particularly easy to compute and
helpful to explain the intuitive idea behind our consistency
requirement.  There is a more general statement of this property to be
satisfied by any background independent spin foam. Namely, spin foam
amplitudes are required to be invariant under the arbitrary
subdivision of their faces. Equivalently, if we deform (by a
piece-wise linear homeomorphism\footnote{For an extensive analysis of
role of piece-wise linear homeomorphisms as opposed to diffeomorphism
as basic symmetry of quantum gravity see \cite{za2}.}) a colored face
by coloring with the same spin adjacent faces in the 2-complex
(previously labeled by the trivial representation) the amplitude
should remain invariant, Figure~\ref{subdi}. We see that with the
normalization found above the Barrett--Crane model satisfies this
necessary condition for background independence as the amplitude of the
composite face $A^{{\rm com}}_f(j)$ can be easily shown to be given by
\begin{equation}
A^{{\rm com}}_f(j)=(2j+1)^{2n_v-2 n_e+2n_f}=(2j+1)^{2 \chi}=A_f(j),
\end{equation}
where $n_e$ and $n_f$ is the number of internal edges and faces of
the composite face. We see that (\ref{amp}) yields
an invariant amplitude.

The analysis here fixes the value of the face amplitude to be
given by the Plancherel measure of the corresponding gauge group.
It is important to notice however that the edge amplitude found here
is only valid for the degenerate situations in which we have
vertex configurations of the form illustrated in Figure \ref{10j}.
These situations are degenerate in the sense that the simplicity
constraints of that reduced $BF$-theory to gravity are trivial. In
the general situation one expects the value of the edge amplitude
to differ from the trivial value obtained here. It is precisely
here where the appropriate Faddeev--Popov factors advocated in
Sections \ref{measure} and \ref{GRe} will play an important role.

The requirements studied here should be met by any theory admitting a
spin foam quantization. In particular it is easy to see that
$BF$-theory in any dimension would satisfy them. This is however to
some extend trivial as $BF$-theory is topological and has a finite
number of degrees of freedom. The following is a simple example of an
(in this sense) background independent spin foam model for a theory
with infinitely many degrees of freedom.

\subsubsection{An anomaly free toy model: a trivial example}\label{te}

In this section we define a spin foam model satisfying the above
minimal requirements of background independence whose elementary
amplitudes are very simple and yet lead to a model that is not
topological. The model is tailored to produce a physically interesting
model that can be thought of as a spin foam quantization of the
Husain--Kucha\v{r} model \cite{husain}.

The action of the Husain--Kucha\v{r} model is given by
\begin{equation}
S[e,A]=\int_M e^i\wedge e^j\wedge F^k(A)\ \epsilon_{ijk},
\end{equation}
where $M$ is a 4-dimensional manifold, $A$ is an $SU(2)$ connection
and $e^i$ is a dreibein field. As it is shown in \cite{husain} one can
use the orientation 4-form, $\epsilon_{abcd}$, to define a densitized
vector field $u^a$ which is orthogonal to the triad.  The previous
action can be written as $SU(2)$-BF-theory in four dimensions, namely
\begin{equation} S[e,A]=\int_M B_k \wedge F^k(A),\end{equation}
supplemented by the constraint $B_{ab}^ku^a=0$ for some given vector
field $u^{a}$.  In addition to the vector field $u^a$ we must provide
an auxiliary space-time foliation which is equivalent to say that we
are given $u_a$ such that $u_a u^a=1$.  If we define the triad $e^a_i$
as $\det(e)\ e^d_i\equiv B_{ab}u_c\epsilon^{abcd}$, where ${\rm
det}(e)$ is defined using $\epsilon_{abc}=\epsilon_{abcd}u^a$, then it
is easy to show that $B^i_{ab}=\epsilon^{i}_{\ jk} e^{j}_ae^{k}_b$
which when replaced in the BF-theory action reproduces the
Husain-Kucha\v{r} action.

\begin{figure}[h]
\centering {\includegraphics[width=4cm]{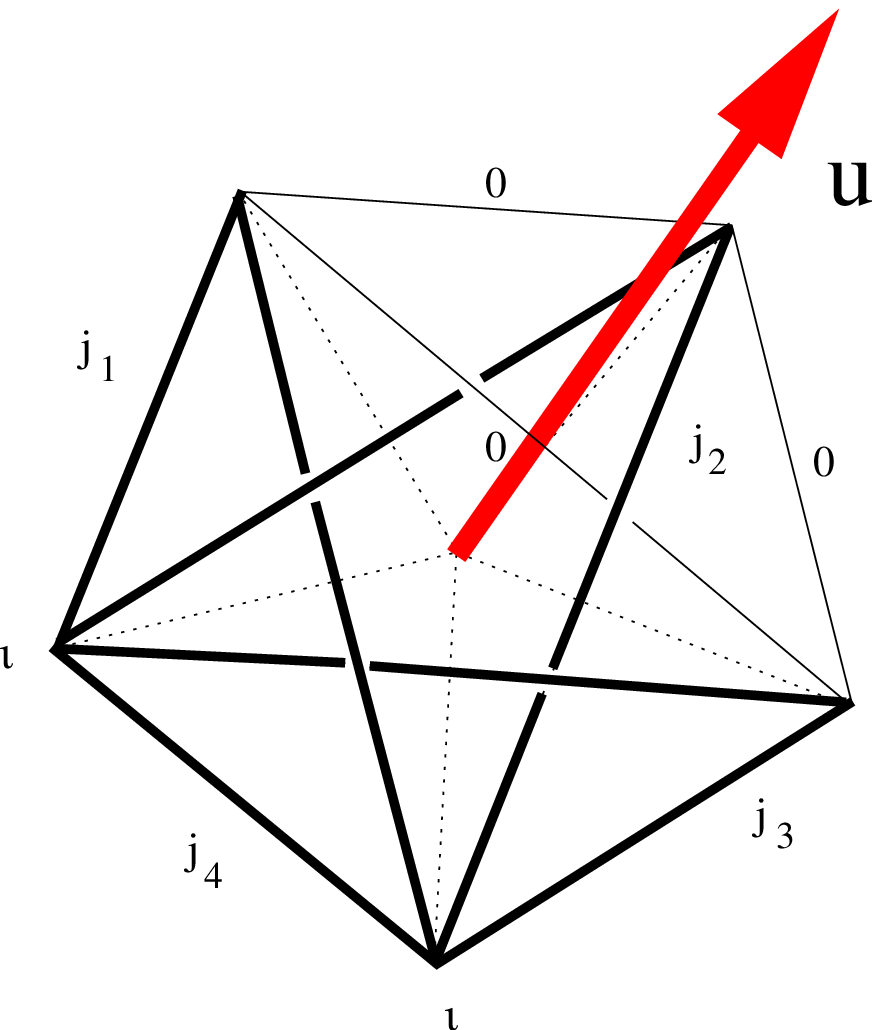}\ \ \ \ \ \
 \ \ \ \ \ \ \ \ \ \ \ \includegraphics[width=4cm]{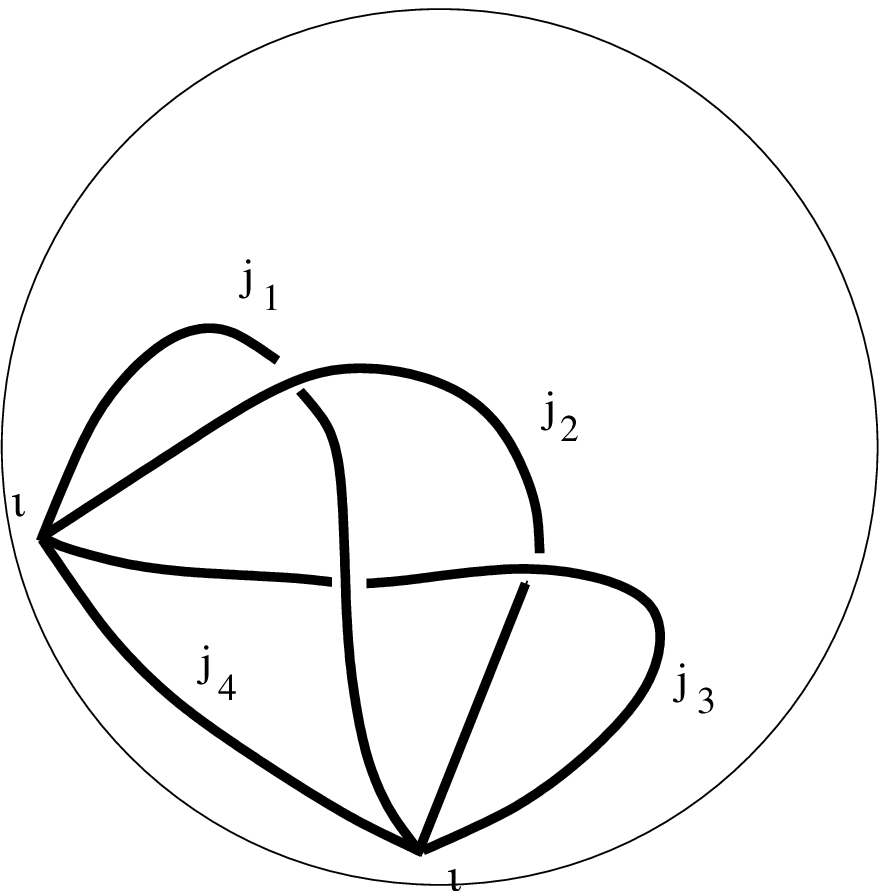}}
 \caption{ On the left: geometric interpretation of the constraints.
 Such spin foam corresponds exactly to an edge carrying an intertwiner
 $\iota$ between four representations $j_1,\cdots j_4$ entering the
 4-simplex and leaving it without change.
% It represents the trivial transition
%from a 4-valent vertex to a 4-valent vertex.
On the right: the continuum representation.
} \label{u}
\end{figure}

For a spin foam quantization we can assume that the model is be
defined on a simplicial decomposition. In this case all vertices in
the dual 2-complex are 5-valent and boundary graphs have 4-valent
nodes. 
%The vertex amplitude of the model is given by the constrained
%$BF$-theory vertex amplitude.  
The standard $BF$-theory amplitude is
the $15j$-symbol constructed with normalized intertwiners (where $j$
are unitary irreducible representations of the corresponding compact
group $G$). Now the constraint $B_{ab}^ku^a=0$ is implemented by
requiring the vertex amplitude to vanish unless at least three
representations labeling the links forming a triangle in the graphical
representation of the $15j$-symbol are trivial, in which case it is
given by the standard $15j$-symbol evaluation.
The idea is illustrated in Figure~\ref{u}. Choosing a triangle on a
4-simplex amounts to choosing a direction $u$ in spacetime. The
components of the $B$-field on the faces that are bounded by this
triangle must vanish according to the constraint $B_{ab}^ku^a=0$ which
is achieved by setting $j=0$ for the corresponding faces.
The resulting spin-network on the boundary of the 4-simplex is simply a
$\theta$-spin-network. Therefore the allowed transitions are very rigid
which implies rather simple configurations.

This model can produce infinite transition amplitudes whenever a
`vacuum' bubble of the kind represented in Figures \ref{vb} or
\ref{vb1} is created.
% Vacuum bubbles appear when we have disconnected
%Wilson loops in intermediate states.  The presence of these
%divergences makes our definition formal at this stage.  
%However the
%simplicity of the fundamental transition amplitudes seems to indicate
%that these divergences could perhaps be controlled in a simple manner.
The model is background independent in the sense above if we set the
face amplitude as the corresponding Plancharel measure of the Lie
group of interest to the power of the Euler characteristic of the
face. For exterior faces one has to modify the amplitude in the usual
way by adapting the definition of the Euler characteristic by counting
by $1/2$ edges with one endpoint on the boundary and $0$ for edges
with two endpoints on the boundary and external faces. 
%It is clear
%that the transition amplitudes produced by the model
%%(leaving aside the issue of regularization) 
%are equal for two spin network states contained in the same
%% (piecewise linear homeomorphism) 
%equivalence class and
%zero otherwise.

The model could be generalized to the case of arbitrary cellular
decompositions, allowing
%. This would allow for the computation of 
arbitrary
%(piecewise linear graph based) 
spin network to spin network
transition amplitudes
%. Namely, we can construct the 
and the construction of a generalized projection operator $P$ to the
physical Hilbert space.  In the continuum limit, e.g.\ defined as in
\cite{za1}, the physical Hilbert space defined by $P$
is much larger than that of $BF$-theory. It would be nice to show its
precise relation to the Hilbert space of spin network states modulo
piecewise linear homeomorphisms---a combinatorial generalization of
the Hilbert space that would correspond to the quantization of the
Husain--Kucha\v{r} model. We emphasize that despite of the simplicity of
the model, it corresponds to a theory with infinitely many degrees of
freedom. This shows that our background independence requirements are
weaker than topological invariance.
%  In order to formalize this one
%would need to deal with the technical issues described above.

In the argument at the beginning of this section we used the existence
of a folliation to show the equivalence of the constraint $BF$-theory
and the action of the Husain-Kucha\v{r} model. Is the model dependent of
this folliation?  The answer is no. At the classical level, the
canonical analysis of the Husain--Kucha\v{r} model uses also a folliation
of the manifold but then it is shown that the equations of motion
imply the independence of such auxiliary structure. In our case our
imposition of the constraints on $BF$-theory selects a direction $u$ at each
4-simplex but this direction is not rigidly specified.  As a
consequence in the spin foam representation one is summing over all
possible $u$'s compatible with the boundary spin networks.
% This means
%that for fixed boundary states we would have different spin foams
%embedded in our discretisation. Many of these will correspond to
%equivalent spin foams in the sense defined at the beginning of this
%section.  In order to define the physical amplitudes one has to
%address the problem of redundancy and sum only over equivalence
%classes of spin foams.

Notice also that the fluctuating character of
$u$ in the state-sum is imposing diffeomorphism invariance on the
boundary. If we compute, e.g., the transition amplitude between
the vacuum (no-spin-network state) and two states
%a state $s_1$ first and then
%the one between the vacuum and $s_2$
in the same (piecewise linear homeomorphism) equivalence class the
contributing configurations of $u$ will be different. Perhaps the
simplest example is already illustrated by Figure~\ref{u}. If we think
of a single 4-simplex as our spacetime then different choices of $u$
correspond to all the different ways of drawing the $\theta$-spin
network on the boundary of a single $4$-simplex.
% These issues could be
%studied in detail in this simple model as they bare some essential
%difficulties of the general problem.

\begin{figure}[h]
\centerline{
\hspace{0.5cm} \(
\begin{array}{c}
\includegraphics[height=2.5cm]{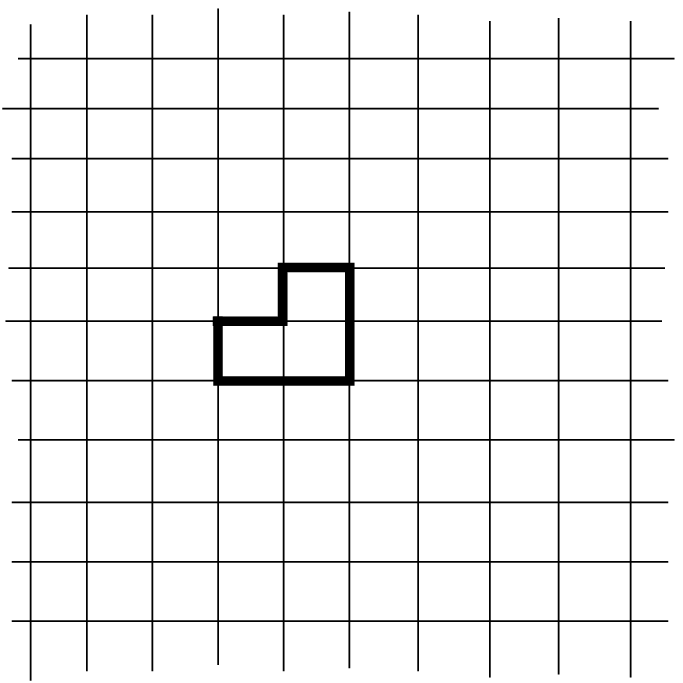}
\end{array}\ \   \phys \ \
\begin{array}{c}
\includegraphics[height=2.5cm]{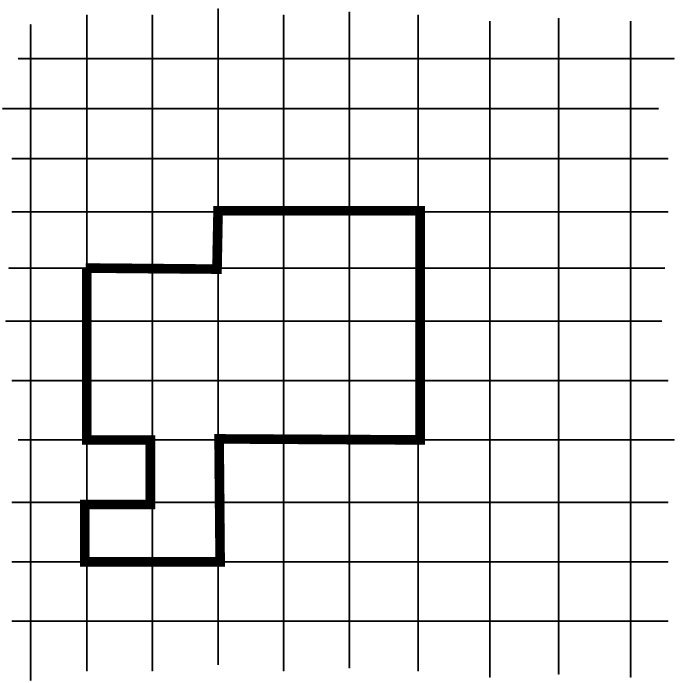}
\end{array}  \ \   \phys \ \
\begin{array}{c}
\includegraphics[height=1.5cm]{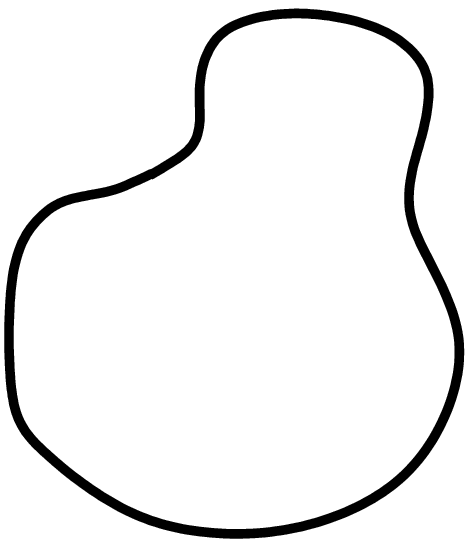}
\end{array}  \) }
\caption{Diagrammatic representation of two distinct Wilson loops that
will be physically equivalent in our toy model. In general any two
spin network states differing by a piecewise linear homeomorphism will
be physically equivalent, i.e, their difference will be in the kernel
of $P$. In the continuum limit one can extend the equivalence to
smooth graphs as the one shown on the right.}
\label{grid}
\end{figure}

Amplitudes in this model are crossing symmetric in the sense of
\cite{reis5,c00}, i.e.\ we are imposing 3-diffeomorphism invariance in
any arbitrary slicing. The purpose of this simple example is to
show how the requirement of 3-diffeomorphism invariance is
directly related to the values of the face and vertex amplitudes
in the special configurations studied in the previous subsection.
In other words, the non triviality of gravity transition
amplitudes (or more precisely its generalized projection operator
$P$) should be encoded in the details of the vertex amplitude for
the configurations that have been avoided {\em de facto} in this
model and otherwise agree with it. From the point of view of the
first part of this paper, the model does not pose any difficulties
since only Lagrange multipliers are constrained to be zero
directly.\footnote{In the canonical treatment of $BF$-theory,
$B_{a0}^i$ are Lagrange multipliers of the constraints
$F_{ab}^i\approx0$. The additional second class constraints for
the Husain--Kucha\v{r} model (vanishing time-components of $B$ and
their momenta) commute with the original first class constraints
of $BF$-theory and have constant bracket among themselves.} There
is thus no additional factor from the constraint algebra and the
lower dimensional amplitudes agree with those of $BF$-theory. This
is in agreement with the minimal requirements used here, which
thus turn out to be sufficient in this case.

\section{Discussion}

The aim of this paper is to point out the need to check the
anomaly problem for spin foam models which in this context means
that lower dimensional simplex amplitudes are not free but
essentially fixed. This is illustrated by the toy model of
Section~\ref{TM} which demonstrates that without the correct
amplitudes non-physical results arise.

In order to find an anomaly-free formulation one has to study the
constraint algebra which in spin foam models for gravity, formulated
as a restricted $BF$-theory, generally involves second class
constraints. For a complicated constraint algebra this requires a
non-trivial function in the measure which has been overlooked
before.
% Furthermore, the constraints, imposed by $\delta$-functions,
%lead to additional functions in the measure which all contribute to
%the face and edge amplitudes.
We have proposed a way to address the issue of computing the
measure in a simplicial theory using the classical phase space
structure.
%starting from the formal path integral measure
%in the continuum.
To get the face and edge amplitude of the discrete
spin foam model we have seen that it is helpful to introduce polar
coordinates for the $B$-field since it provides a direct link to the
spin parameters of the state sum. Explicit calculations require
several integrations which, if not possible to be done explicitly, are
well-suited for a stationary phase approximation.
% One has to be
%careful, however, because the integrand can be singular. 
A detailed
study is necessary in order to be able to judge if amplitudes of a
particular model would be finite. 
%We hope to come back to this issue
%for the case of gravity in a future publication.

Even though we have not explicitly computed the anomaly free
measure for spin foam models of gravity we have pointed out that
some minimal requirements from the related condition of
background independence can be imposed that severely restrict the
value of the face and edge amplitudes. The restrictions imposed by
background independence in the way presented here seem to imply
that bubble divergences in spin foam models for gravity are linked
with the gauge action of diffeomorphisms, and
%. We have shown that these restrictions 
rule out some proposals in the literature. In particular, without
modification of the singular edge amplitude
(associated with edges bounded by less than four faces), the finite
normalizations for both the Riemannian and Lorentzian Barrett--Crane
model proposed in \cite{a10, a9} are to be regarded as formulations
where the diffeomorphism gauge symmetry has been broken by an
anomalous path integral measure.\footnote{This (anomalous) finite
normalization of the Barrett--Crane model was naturally obtained in
the context of the group field theory (GFT) formulation of the
model. One could modify the amplitudes of singular edges and make the
model anomaly free in the sense described here.  In this way there
will be divergences but of a rather simple kind (only isolated vacuum
bubble will diverge and could be easily renormalized). However, the
naturality argument in relation to a GFT formulation would not stand
in this case.} Other anomalous formulations are: the new normalization
of the Barrett--Crane model proposed in \cite{baez1} to improve the
convergence properties of the previous model and Model A in
\cite{fre2}. This seems to severely limit the physical relevance
of such proposals. Model B in \cite{fre2} is the only
normalization of the Barrett--Crane model which satisfies the
minimal requirements of background independence presented here
and is the one naturally arising in the quantization of
Plebanski's $Spin(4)$ formulation of gravity presented in
\cite{a16}. However, a non trivial modification of the edge
amplitude for generic edges should appear due to the contribution
of the simplicity constraints as argued in Section~\ref{measure}.

The value of the correct edge amplitude can be determined if we
understand the canonical algebra of simplicity constraints so that the
appropriate determinant as in (\ref{Zsec}) is included. This is
a complicated issue as it might require the understanding of the
canonical formulation of the simplicial model. Perhaps the ideas of
Gambini and Pullin in the context of their {\em consistent
discretization} formulation might shed some new light on this
issue. Unfortunately the canonical formulation of discrete theories
seems rather complicated in the case of Plebanski's formulation at
this stage.
An interesting alternative procedure to deal with constrained systems
is the projection-operator approach \cite{Klau1,Klau2,Klau3}. Since
this method allows to deal with first and second class constraints on
an equal footing, it may be possible to sidestep some of the
difficulties mentioned above.

The well understood results in 3-dimensional gravity \cite{fre7}
and those of Section \ref{DC} suggest that the discretization does
not completely fix the diffeomorphism gauge transformations as it
is usually assumed. In the case of the Barrett--Crane model (any
other model could be analyzed in this way) we have shown that the
minimal requirements for background independence already imply the
presence of certain bubble divergences that have no physical
content. For instance, even in our toy (Husain--Kucha\v{r}) spin foam
model divergences can not be avoided without some extra
manipulation.

These divergences will be present in any spin foam model for a
background independent theory and the way to deal with them is by
appropriate gauge fixing conditions. 
%This is in fact possible in the case of three dimensional gravity. 
%In four dimensions one
For this, one needs to understand in a precise manner the action of
diffeomorphisms in the context of the simplicial models. Our
minimal requirements of background independence of Section
\ref{DC} are closely related to the action of 3-diffeomorphisms in
loop quantum gravity (it is tempting to think that due to the fact
that our requirements hold for any `slicing' of the spin foam one
is imposing a `bit' of 4-diffeos in this picture). In this sense,
what is left to understand is the old questions: where is the
remnant gauge transformation encoded in the action of the
Hamiltonian constraint in the canonical framework?, and can we
expect to be able to separate the physical dynamics from the gauge
evolution by a closer analysis of the vertex amplitude?

%Although one would need to understand the technical details of the
%generalization to 4-dimensional $BF$-theory, the divergence structure of
%the Crane-Yetter model suggests that the Freidel--Louapre analysis should
%go through in this case. Going to 
In the Barrett--Crane model we
constrain the $B$ field to be given by a simple bivector field
derived from a tetrad. At the spin foam level, $Spin(4)$
representations are constrained to simple representations
$j\otimes j^*$. The essential questions are: what part of
`translational' gauge freedom (\ref{topo}) in the $B$'s of $Spin(4)$
$BF$-theory remains after the implementation of the simplicity
constraints? Namely: Is this gauge symmetry remnant fully encoded in
the equivalence class of spin foams defined by Baez?, or is there also
a symmetry that can change the values of the representation
labels as in $BF$-theory?

In the first scenario the gauge divergence structure of the model does
not seem problematic. We have seen in Section \ref{measure} that in
addition to the Faddeev--Popov factor coming from gauge fixing first
class constraints the measure should be modified when implementing the
simplicity constraints. From the analysis of Section \ref{DC} we
conclude that this modification should involve the value of the edge
amplitude for non-singular edges. The implementation of the simplicity
constraints could modify the amplitudes in a way that would make
non-singular bubble amplitudes finite. One could think of the non
trivial damping edge amplitude of the finite model \cite{a7} as
arising in this way for non-singular edges. These factors must not
arise for singular edges bounding only two non trivial faces as our
consistency argument fixes the edge amplitude to unity in these cases.
In this scenario the divergence structure of the (non-gauge fixed)
amplitudes would be similar to that of our toy model of
Section~\ref{DC}: only the vacuum bubbles as the ones represented in
Figures \ref{vb} or \ref{vb1} would be the divergent contributions to
the amplitudes and can be easily regularized.

In the second scenario, it is appealing to think that the sum over
representations `flowing' inside a bubble is un-physical and its
contribution corresponds to the diffeomorphism gauge volume. Gauge
fixing will correspond to dropping these redundant sums and replace
them by the appropriate Faddeev--Popov determinants of Section
\ref{measure}. In the case of $BF$-theory topological invariance follows
from the triviality of these Faddeev--Popov determinants.
% (they are all equal
%to one due to the presence of the $\delta$-functions that set the
%curvature to zero \cite{fre7}). 
In the case of gravity these factors should
depend in a non-trivial fashion on the spins as the correct model must
contain local excitations.

\ack

We are grateful to A Ashtekar, D Christensen, L Freidel,
R Gambini, J Klauder, J Pullin, C Rovelli and T Thiemann for
discussions and to J Klauder for hospitality to one of us (MB). We
also thank the Erwin Schr\"odinger Institute, Vienna, where part of
the work has been done at the workshop ``Quantum field theory in
curved spacetime''. This work was supported in part by NSF grant
PHY00-90091 and the Eberly research funds of Penn State.

\section*{Appendix: A degenerate sector of gravity}

In this appendix we illustrate how the study of the integrals that
define the amplitude of spin foam models obtained by discretizing a
continuous action can be used to find the correct normalization of the
spin foam measure (in the large spin limit).  The example presented
here is a little bit more involved than the one of
Subsection~\ref{BFp} and is closely related to the Barrett-Crane
model.  Even though the analog of simplicity constraints is present in
this example, it should be pointed out that our analysis here has an
important limitation: we are not including the determinant factors
appearing in (\ref{Zsec}) produced by the implementation of second
class constraints in the path integral.  This example is however meant
as a computation that illustrates the difficulties involved in the
analysis. We have kept this calculation as an appendix hoping that the
technique can be useful for further developments.

A degenerate sector of Euclidean gravity can be obtained by
introducing the additional constraint \cite{reis0}
%\begin{equation}\label{degconst}
$B^L_i=V_i^jB^R_j$
%\end{equation}
into $SO(4)$ $BF$-theory where $B^L$ and $B^R$ are $su(2)$
valued components of the $so(4)$ valued $B$ according to the
decomposition $SO(4)\cong SU(2)\times SU(2)$, and $V$ is an
$SO(3)$-matrix. The new action then is
\[
 S=\int \left(B_i^L\wedge F_i^L+B_i^R\wedge F_i^R+\lambda^i\wedge
 (B_i^L-V_i^jB_j^R)\right)
\]
where $F_i^L$ and $F_i^R$ denote the curvatures of the left and
right component of the connection, respectively. In this form,
$\lambda^i_{ab}$ and $V_i^j$ appear as new Lagrange multipliers in
addition to the multipliers $B_{0a}^{L/R,i}$ and $A_0^{L/R,i}$ of
the original $BF$-theory. Only the multipliers $B_{0a}^{L,i}$ and
some components of $\lambda_{ab}^i$ are restricted by the new
constraints and in order to obtain a constrained system on a
symplectic phase space we have to add their momenta to the
canonical variables together with constraints requiring them to be
zero. 
%The algebra of all the constraints will determine the
%measure we have to use in the path integral. 
For a discussion of the finiteness of the resulting model it is most
interesting to see whether or not additional factors in the measure
depend on components of $B$ with a non-trivial scaling behavior since
this would affect the large-$j$ behavior of face amplitudes. This has
to be expected here because varying $V_i^j$ yields a constraint which
restricts components of the multipliers $\lambda_{ab}^i$ and is linear
in $B$. Since we had to add momenta of these components of
$\lambda_{ab}^i$, the constraint algebra will contribute positive
powers of $B$ to the measure which would enhance a divergence.  Here,
however, we do not discuss the constraints in detail but rather ignore
the additional factor and derive the large-$j$ behavior of the face
amplitude for the naive spin foam quantization with trivial multiplier
measure. While the resulting amplitude would not be correct, our aim
here is solely to compare it with the result of \cite{a16} where this
factor has been ignored, too.

To include the additional constraint into a spin foam quantization
we impose it face-wise such that any face $f$ carries a
matrix $V_f$. This gives the state sum
\begin{eqnarray*}
\fl Z = \int\prod_e\d^3g_e^L\d^3g_e^R
 \prod_f\d^3B_f^R\d^3V_f \exp\left(\i\tr(B_f^R(U_f^R+U_f^LV_f))\right)\\
 \fl= \int\prod_e\d^3g_e^L\d^3g_e^R
 \prod_f\d^3B_f^L\d^3B_f^R\d^3V_f\d^3\lambda_f
  \exp\left(\i\tr(B_f^LU_f^L+B_f^RU_f^R+
 \lambda_f(B_f^L-V_fB_f^R))\right)
\end{eqnarray*}
where $\lambda_f$ are Lagrange multipliers and the edge holonomies
$(g_e^L,g_e^R)$ form the holonomies $U_f^L$ and $U_f^R$ along closed
loops. Integrating over $\lambda_f$ yields $\delta$-functions which
will be solved after introducing polar coordinates
$(r^{L/R},\vt^{L/R},\vp^{L/R})$ for $B^{L/R}$ and Euler angles
$(\psi,\theta,\phi)$ for $V$. The $\delta$-functions then imply
$r_f^L=r_f^R$ for all faces $f$, and $\vt^L$ and $\vp^L$ will be given
as functions $\vt^L=F(\vt^R,\psi-\vp^R,\theta)$ according to
$\cos\vt^L=\sin\theta\sin\vt^R\sin(\psi-\vp^R)+\cos\theta\cos\vt^R$
and $\vp^L=G(\vt^R,\vp^R,\psi,\theta,\phi)$. Choosing again a
gauge for $U^{L/R}$ without loss of generality, we obtain
\begin{eqnarray*}
\fl Z = \int\prod_e\d^3g_e^L\d^3g_e^R
 \prod_f\d^3B_f^L\d^3B_f^R\d^3V_f \delta^3(B_f^L-V_fB_f^R)
 \exp\left(\i\tr(B_f^RU_f^R+B_f^LU_f^L))\right)\\
 \fl = \int\prod_e\d^3g_e^L\d^3g_e^R \prod_f\d
 r_f^L\d\vt_f^L\d\vp_f^L (r_f^L)^2\sin\vt_f^L \d
 r_f^R\d\vt_f^R\d\vp_f^R (r_f^R)^2\sin\vt_f^R
 \d\psi_f\d\theta_f\d\phi_f\sin\theta_f\\
 \fl \quad\times((r_f^L)^2\sin\vt_f^L)^{-1}\delta(r_f^L-r_f^R)
 \delta(\vt_f^L-F(\vt_f^R,\psi_f-\vp_f^R,\theta_f))
 \delta(\vp_f^L-G(\vt_f^R,\vp_f^R,\psi_f,\theta_f,\phi_f))\\
 \fl\quad\times\exp\left(-\i(r_f^L\cos\vt_f^L\sin\case{c_f^L}{2}+
 r_f^R\cos\vt_f^R\sin\case{c_f^R}{2})\right)\\
 \fl= \int\prod_e\d^3g_e^L\d^3g_e^R \prod_f
 \d r_f^R\d\vt_f^R\d\vp_f^R (r_f^R)^2\sin\vt_f^R
 \d\psi_f\d\theta_f\d\phi_f\sin\theta_f\\
 \fl\times\exp\left(-\i r_f^R(\cos
 F(\vt_f^R,\psi_f-\vp_f^R,\theta_f)\sin\case{c_f^L}{2}+
 \cos\vt_f^R\sin\case{c_f^R}{2})\right)\,.
\end{eqnarray*}
We now have to perform the angle integrations, which is effectively a
four-dimensional integral since the $\phi$- and
$\psi+\vp^R$-integrations are trivial, in order to find the face
amplitude after discretizing $r^R$. One can use the saddle point
approximation, but has to be careful because the integrand has zeroes
due to the sines.
Disregarding the zeroes, one would expect the asymptotic behavior of
the angle integral to be $(r^R)^{-2}$ because every one-dimensional
integration contributes $r^{-\frac{1}{2}}$ in the standard stationary
phase result. This factor would cancel the $(r^R)^2$ from the measure
and thus yield $1$ as the large-$j$ behavior of the face amplitude. In
\cite{a16} a face amplitude $A_f(j)=1$ has been obtained for this
model by different means, however for a different discretization of
the action which is not accessible to our methods.  Nevertheless,
there is agreement provided that the modified discretization in
\cite{a16} leads to the standard stationary phase result. An
analysis similar to that of Section~\ref{DC} would lead to the
conclusion that such a face amplitude is in conflict with background
independence. This should not be surprising as there is no reason for
amplitudes to be anomaly free when we do not include the correct
measure in the presence of second class constraints. At present we do
not see whether doing so in this case would lead to amplitudes in
agreement with background independence.

\end{document}